\documentstyle[12pt,fleqn,epsfig,rotate]{article}
\begin{document}

\title{The Pt isotopes: comparing the Interacting Boson Model with
Configuration Mixing and the Extended Consistent-Q formalism}
\vspace{2cm}

\author{J.E.~Garc\'{\i}a-Ramos $^1$ and K.~Heyde $^2$\\~\\
$^1$ {\small Departamento de  F\'{\i}sica Aplicada,} \\
{\small Universidad de Huelva, 21071 Huelva, Spain}\\ \\
$^2$ {\small Department of Subatomic and Radiation Physics,}\\ 
 {\small Proeftuinstraat, 86 B-9000 Gent, Belgium}}
\maketitle

\begin{abstract} 
The role of intruder configurations in the description of energy
spectra  and $B(E2)$ values
in the Pt region
is analyzed.
In particular, we study the differences between
Interacting Boson Model calculations with or without the inclusion of
intruder states in the even Pt nuclei $^{172-194}$Pt. 
As a result, it shows that for the description of a subset of the
existing experimental data, {\it i.e.}, 
energy spectra and absolute $B(E2)$ values up to an excitation energy of about $1.5$ MeV, both
approaches seem to be equally valid. We explain these similarities between both model spaces
through an appropriate mapping.
We point out the need for a more extensive comparison, encompassing a data set as broad (and
complete) as possible to confront with both theoretical approaches  in order to test the detailed structure
of the nuclear wave functions.
\end{abstract}



\section{Introduction}
\label{sec-intro}
The appearance of shape coexistence in nuclei has attracted a lot of
attention in recent decades \cite{hey83,wood92,jul01} and
compelling evidence has been obtained, in particular, at and very near
to proton or neutron closed shells.  The Pb region takes
a prominent position because in addition to specific shell-model excitations
established close to the $N=126$ neutron shell closure, collective excitations have
been observed for the neutron-deficient nuclei \cite{andrei00}. The
review paper of Julin {\it et al.} \cite{jul01} gives an extensive overview
of the rich variety in nuclear excitation modes for both the Pb, Hg, Pt
and, Po, Rn nuclei. Since the publication of that paper, however,
many experiments have been performed highlighting properties of
excited band structures and lifetimes for the Pb nuclei 
\cite{dracou03,vanvel03,dewald03,reviol03,page03,dracou04,pakarin07,grahn06,grahn08} 
and nearby nuclei, also illuminating
the underlying mechanisms that are at the origin of the formation of
collective excitations in the neutron-deficient nuclei near to the $Z=82$ proton closed shell.

This mass region has been studied theoretically in an extensive way.
Early calculations using a deformed Woods-Saxon potential in order to
explore the nuclear energy surfaces as a function of the quadrupole deformation
variables \cite{may77,bengt87,bengt89,naza93} showed a consistent
picture pointing out the presence of oblate and prolate energy minima.
More recently, mean-field calculations going beyond the static part,
including dynamical effects using the Generator Coordinate Method
(GCM) \cite{bender03}, either starting from Skyrme functionals
\cite{grahn08,duguet03,smirnova03,bender04}, or using the Gogny D1S
parametrization \cite{girod89,chasman01, egido04,rodri04,sarri08} have put the
former calculations on firm ground and have also given rise to
detailed information of the collective bands observed in
neutron-deficient nuclei around the $Z=82$ closed proton shell. Here, we should
also mention attempts to study shape transitions in the Os and Pt nuclei
within the relativistic mean field (RMF) approach \cite{fossion06}. 

>From a microscopic shell-model approach, the hope to treat on equal footing the
large open neutron shell from $N=126$ down to and beyond the mid-shell
$N=104$ region, with valence protons in the Pt, Hg, Po, and Rn
nuclei, and even
including proton multi-particle multi-hole (mp-nh) excitations across
the $Z=82$ shell closure, is far beyond present computational possibilities.
The truncation of the model space, however, by concentrating on nucleon pair
modes (mainly $0^+$ and $2^+$ coupled pairs, to be treated as bosons
within the Interacting Boson Approximation (IBM) \cite{iach87}),
has made the calculations feasible, even including pair excitations
across the $Z=82$ shell closure \cite{duval82} in the Pb region. More
in particular, the Pb nuclei have been extensively studied giving rise
to bands with varying collectivity depending on the nature of the
excitations treated in the model space
\cite{pakarin07,hey87,hey91,hey94,coster96,oros99,fossion03,helle05,helle08}.

In view of the fact that near the mid-shell point of the valence
neutron major shell $82-126$ ($N=104$) clear-cut examples (see the
discussion before) of coexisting collective bands have been observed
in both the Pb ($Z=82$) and Hg ($Z=80$) isotopes, it is of the utmost
importance to study the propagation of these coexisting structures as
one gradually moves away from the $Z=82$ proton closed
shell. Therefore, detailed studies for both the Pt, Os,... isotopes
($Z=78, 76$,...)  and the Po, Rn,... isotopes ($Z=84, 86$,...) are of
major importance in order to explore the evolution from coexisting
spherical and deformed structures at the closed shell towards the
onset of normal, open-shell deformation. In order to gain insight into this transition, one
should, in particular, study those observables that are sensitive to
specific mp-nh excitations such as the appearance of distinct
collective bands making use, in particular, of Coulomb excitation with radioactive beams,
$\alpha$-decay hindrance factors, isotopic shifts, E0 decay
properties, g-factors.  Moreover, data on low-lying excited states in
the adjacent odd-proton mass Au, Tl, Bi,...~isotopes as well as in the
odd-neutron mass Pt, Hg, Pb, Po, Rn isotopes should allow to explore
the importance of particular single-particle excitations in this mass
region.  The Pt nuclei form a most important set of isotopes in order
to study the above question.

In the present paper, we reanalyze the Pt nuclei, motivated by the above
questions and recent
IBM calculations \cite{cutcham05, cutcham05a} without explicitly
including intruder excitations into the model space, considering the 4
proton holes and the number of valence neutrons with the boson
model approximation. Before, it has been suggested by Wood
\cite{wood81,wood82} that proton 6h-2p configurations  must be
considered, besides the regular proton 4h configurations, in order to
describe the experimental data in the Pt and adjacent nuclei, in
particular the sudden lowering of the first excited $0^+_2$ down to mass
number $A=186$ associated with a specific change in the excitation
energy for the first excited $2^+_1$ state in the same mass interval.
Thus, it becomes clear that the Pt nuclei form a most important series of isotopes in order to study
the above question.
More specifically, within the IBM configuration-mixing approach \cite{duval82}
(IBM-CM for short), calculations have been carried out for the Pt nuclei
by Harder {\it et al.} \cite{harder97} and by King {\it et al.} \cite{king98} describing the low-lying
excited states in the $^{168-196}$Pt nuclei. In the latter studies, also
$g(2^+_1)$ factors and isotopic shifts were calculated and shown to be in good
agreement with the known data \cite{stuchbery96,hilberath92}. However, no $B(E2)$
values, which constitute an important test, were calculated.

We should mention that the even-even Pt around mass number $A \approx 190$ have been studied before
in the framework of the proton-neutron interacting boson model by Bijker {\it et al.}~\cite{bijker80}. 
They calculated a large number of observables such as the energy spectra, various 
ratios of reduced $E2$ transition probabilities, absolute $B(E2)$ values and quadrupole moments,
two-nucleon transfer reaction intensities, isotopic and isomeric shifts and E0 transition rates. 
In view of the fact that this study dates back to 1980, an extensive
set of new data have become available since. This warrants an in-depth
study of the Pt nuclei, spanning the region from the heavier Pt nuclei (around $A \approx 196,198$) into
the lightest Pt nuclei near A $\approx$ 172.

In the present study, we show that for the excited states below an
excitation energy of $E_x \approx 1.5$ MeV, the IBM calculations with
or without including explicitly particle-hole excitations 
reproduce equally well the excitation energies and absolute $B(E2)$
values of known states in the mass region $172\leq A \leq 194$. 
How then can we interpret a possible equivalence of the energies, 
and the $B(E2)$ values for a limited number of
states, below $1.5$ MeV between both calculations? 
This excitation energy approximately marks the start of a region in which a unique comparison
between a specific experimental level and a particular level in the IBM calculations becomes difficult
to be made.
First of all, we point
out that even though the Hilbert spaces of both models are largely
different, due to the limited number of states below $1.5$ MeV, it is
not possible to discriminate between the models considering only 
excitation energies and $B(E2)$ values. Secondly, we point out that
the difference in the Hamiltonians can result from a renormalization
of the interaction. 
We also show that, moving towards higher-lying
levels, clear differences appear between the model spaces.
Moreover, we explore theoretical possibilities to understand the fact that excitation energies
and electromagnetic quadrupole properties ($B(E2)$ values and quadrupole moments) exhibit
a very strong ressemblance, although originating from highly different wave functions.
In a forthcoming paper, we shall present  
the results of a comparison covering an as complete as possible data set (encompassing
also  $\alpha$-decay hindrance 
factors, $g(2^+_1)$ factors, isotopic shifts, E0 decay properties, and also 
properties of the energy spectra of adjacent odd-mass Pt and Au nuclei) with the two theoretical
approaches, {\it i.e.,} considering a reduced model space versus an enlarged
model space including particle-hole excitations, within the Interacting Boson
Model context.
\begin{figure}[hbt]
  \centering
\hspace*{-3cm}  \epsfig{file=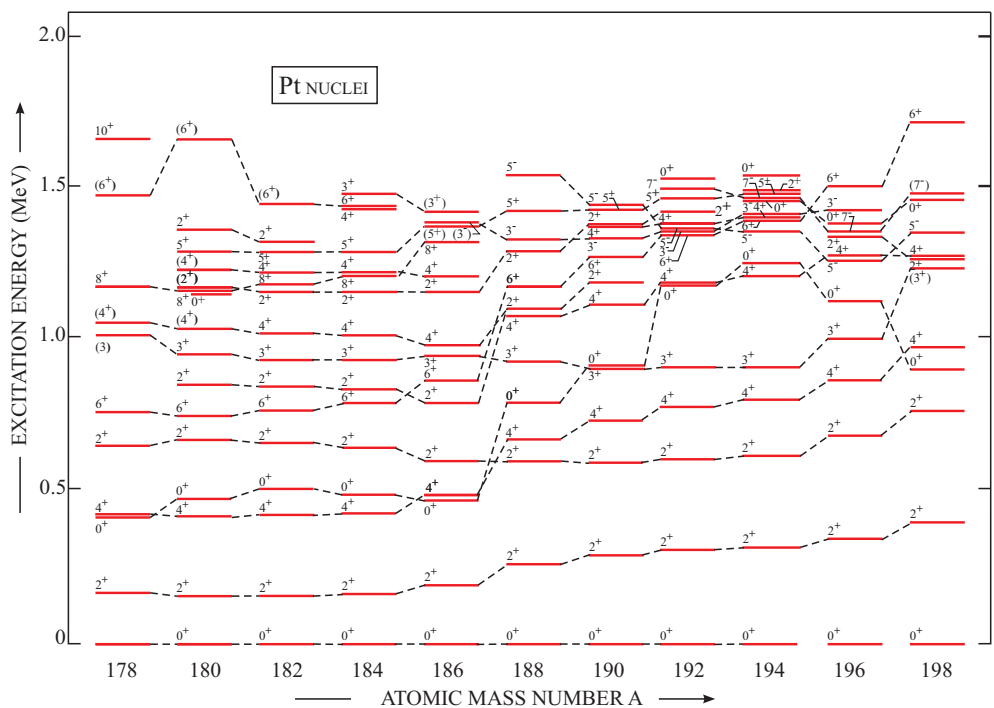,width=17cm} 
\vspace*{-2cm}
  \caption{Energy level systematics for Pt isotopes.
Only levels up to $E_x \sim$ 1.5 MeV are shown.}
  \label{fig-system-wood}
\end{figure}

\section{The experimental data in the even-even Pt nuclei}
\label{sec-expe}
The even-even Pt nuclei, in particular the region of neutron-deficient
isotopes situated around and even beyond $N=104$ neutron number ($A=182$),
spanning the mass interval  
$172\leq A \leq 194$, have been extensively studied over the
last decades. Information was taken from the appropriate
Nuclear Data Sheets covering the above Pt mass region for $A=172$
\cite{Sing95}, $A=174$ 
\cite{Brow99}, $A=176$ \cite{Basu06}, $A=178$ \cite{Brow94}, $A=180$ \cite{Wu03}, 
$A=182$ \cite{Sing95b}, $A=184$ \cite{Fire89}, $A=186$ \cite{Bagl03},
$A=188$ \cite{Sing02},  $A=190$ \cite{Sing03}, 
$A=192$ \cite{Bagl98} and, $A=194$ \cite{Sing06}.
Moreover, we also incorporate data on those even-even Pt nuclei that have been published
(see below) since the latest updates of the corresponding Nuclear Data Sheets.

The very light Pt nuclei with $A=168,170$ were studied by King {\it et
  al.}~\cite{king98} using $\alpha$-decay recoil tagging gamma spectroscopy. 
In the case of $^{170}$Pt, a ground-state band up to spin $(8^+)$ was established.
Cederwall {\it et al.}~\cite{cederwal98} obtained yrast bands for both $^{171,172}$Pt up
to spin $(25/2^+)$ and $(8^+)$, respectively. 
Seweryniak {\it et al.}~\cite{sewer98} also studied the
$^{170,171,172}$Pt nuclei and in the case of $A=172$ could extend the
yrast band structure up to spin $(14^+)$, using recoil decay tagging.
More recently, Joss {\it et al.}~\cite{joss06} studied the
neutron-deficient $^{169-173}$Pt nuclei 
obtaining more detailed information on
the yrast bands for mass $A=170$ and $A=172$, and obtaining first information
on side bands. 
Extra information on $^{174}$Pt \cite{goon04} and $^{178}$Pt \cite{kondev00} has been
obtained 
making use of the Gammasphere array, in particular, extending the yrast band up 
to spins $(26^+)$ and $(24^+)$, respectively.
The yrast band in $^{178}$Pt was extended by Soramel {\it et al.}~\cite{soramel99} up to spin
$(18^+)$ using fusion-evaporation reactions. 
Popescu {\it et al.}~\cite{popescu97}, have studied
the high-spin states (up to spin $26^+$) as well as the band structures in $^{182}$Pt.
A very detailed study was performed by Davidson {\it et
  al.}~\cite{davidson99} 
via $\beta^+/EC$ decay using a gamma array
to obtain both, yrast and non-yrast structures covering the
$^{176-182}$Pt nuclei. In particular, the data for $A=178$ and $A=182$
extend the information compiled 
in the corresponding Nuclear Data Sheets \cite{Brow94} and
\cite{Sing95b}, respectively.   
In $^{180}$Pt, lifetime measurements have been performed that allow to extract the 
$B(E2;4^+_1 \rightarrow 2^+_1)$ value. The experiment of De Voigt {\it
  et al.}~\cite{devoigt90} results
in a value of 140$\pm$30 W.u. A more recent experiment by Williams {\it et al.}~\cite{williams06} 
has resulted in a much larger $B(E2)$ value of 260$\pm$32 W.u.~thus giving rise to non-overlapping data.
The mass $A=184$ Pt nucleus was studied by Xu {\it et al.}~\cite{xu92}
in great detail, making use of the $\beta$ decay of $^{184}$Au.  
Using both conversion electron and $\gamma$-ray spectroscopy as well as $\gamma$-ray 
angular distribution measurements on low-temperature oriented (LTNO) $^{184}$Au nuclei, E0 
$\rho^2$ values could be extracted demonstrating coexisting K=0 and K=2 bands.  
The high-spin states in $^{192}$Pt
nucleus have been studied quite recently by Oktem {\it et al.}~\cite{oktem07}, 
going up to spin $(20^+)$, and  by  McCutchan {\it et al.}~\cite{cutcham08}
studying low-spin states in $^{192}$Pt populated in $\beta^+$/EC decay of $^{192}$Au(1$^-$).
Coulomb excitation experiments on $^{194}$Pt by Wu {\it et al.}~\cite{Wu96} have resulted in
an extensive set of reduced E2 matrix elements for the ground-band, the K=2 band (encompassing
both transition and diagonal matrix elements) as well as for E2 transitions decaying from the
$0^+_{2,4}$ states.

The experimental energy systematics for the Pt nuclei, including
the region of interest which is mainly situated around the neutron
$N=104$ mid-shell number, is shown in Fig.~\ref{fig-system-wood} and
spans the interval $178 \leq A \leq 198$. The systematics is limited
to levels with an excitation energy up to $E_x \approx 1.5$ MeV 
(a limited number of higher-spin states are given beyond this cut-off)
and will serve as a ``basis'' to compare with the calculations described
in section \ref{sec-comparing}. This figure is based on the information available in the 
appropriate Nuclear Data Sheets, enlarged or improved including the more recent papers discussing these
particular even-even Pt nuclei. Figure \ref{fig-system-wood}
is characterized by a sudden drop of the first excited $0^+_2$ state, down to mass $A=186$,
remaining essentially constant for lower mass numbers (down to mass $A=178$). The energy spectrum remains
remarkably constant in the mass interval $178 \leq A \leq 186$, 
contrasting with a different structure that shows a number of up sloping levels
with increasing mass number $A$ beyond mass $A=186$.  
\begin{figure}[hbt]
  \centering
  \epsfig{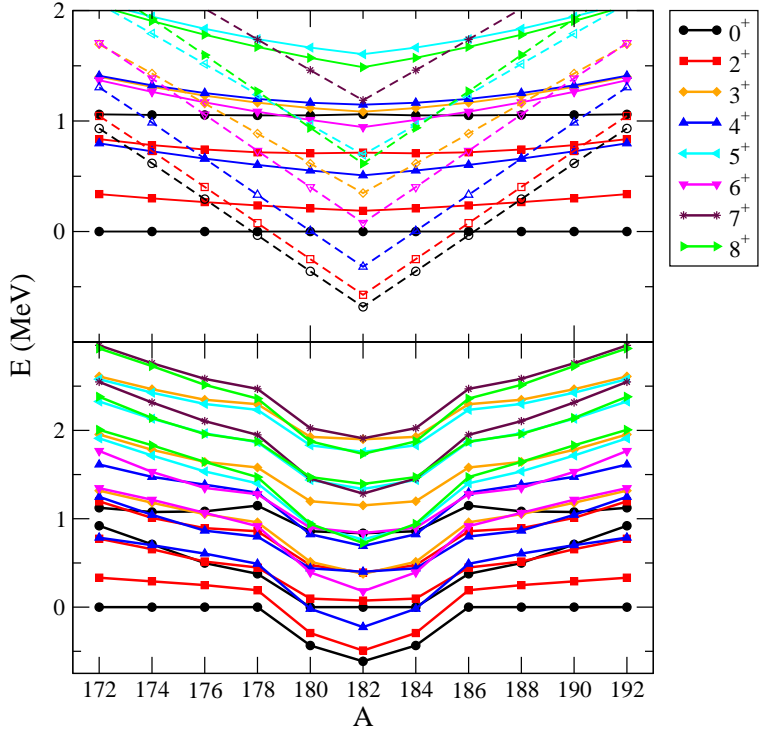}
  \caption{Excitation energies resulting from an IBM-CM Hamiltonian, using the
    parameters taken from Harder {\it et al.} \cite{harder97}. Full lines with closed
    symbols correspond to the
    regular states, while dashed lines with open symbols correspond to the intruder
    states. 
    Upper panel: calculation without the
    mixing term, relative to the first regular $0^+$ configuration. 
    Lower panel: calculation including the mixing term with a strength equal to $50$ keV, 
    relative to  the $0^+$ state with the highest percentage of the regular N-boson subspace.}
  \label{fig-piet-mix}
\end{figure}

\section{Comparing the Interacting Boson Model with configuration
  mixing and the extended Consis\-tent-Q formalism}
\label{sec-comparing}
\subsection{The formalism}
\label{sec-formalism}

In the present study we compare calculations carried out
within the context of the IBM, incorporating 2p-2h excitations across
the $Z=82$ proton closed shell into the model space (also called
IBM-configuration mixing, or IBM-CM as a shorthand notation), with
recent studies using the standard IBM in which excitations across the $Z=82$
core are not included explicitly.
\begin{figure}[hbt]
  \centering
  \epsfig{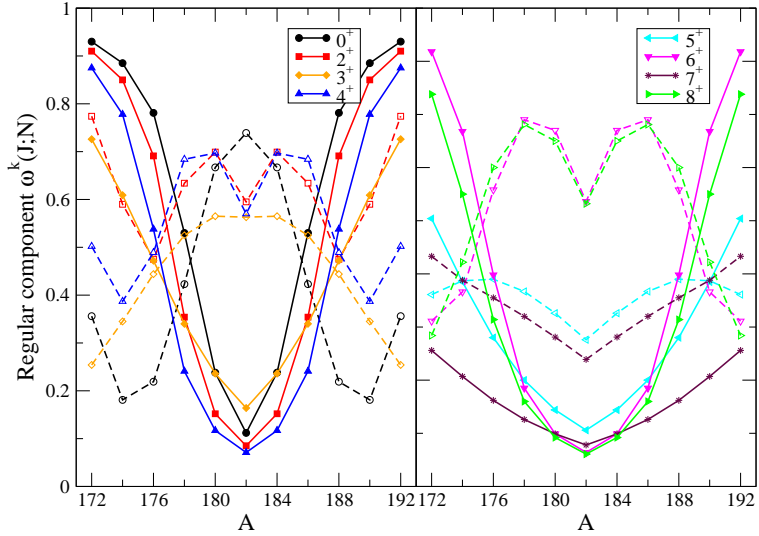}
  \caption {The regular configuration content, expressed by the weight 
    $w^k(J,N)$ (see text), for the two lowest-lying 
    states $(k=1,2)$ 
    (full lines with
    closed symbols are used for the first state while dashed lines with open
    symbols are used for the second state)
    for each of the angular momenta $0^+,2^+,3^+,4^+,5^+,6^+,7^+,8^+$,
    corresponding to the lower panel of Fig.~\ref{fig-piet-mix}.}
  \label{fig-piet-mix-wf}
\end{figure}

The IBM-CM allows the simultaneous treatment and mixing of several
boson configurations which correspond to different particle--hole
(p--h) shell  excitations \cite{duval82}. On the basis of
intruder spin symmetry \cite{coster96,hey92}, no distinction is made
between particle and hole bosons. Hence, the model space which
includes the regular proton 4h configurations and a number of valence
neutrons outside of the $N=82$ closed shell (corresponding to the
standard IBM treatment for the Pt even-even nuclei) as well as the
proton 6h-2p configurations and the same number of valence neutrons
corresponds to a $[N]\oplus[N+2]$ boson space ($N$, being the number of
active protons, counting both, proton holes and particles, plus the
number of valence neutrons outside the $N=126$ or the $N=82$ closed shell
(depending on the nearest closed shell),
divided by $2$ as the boson number). Consequently, the Hamiltonian for
two-configuration mixing can be written as
\begin{equation}
  \hat{H}=\hat{P}^{\dag}_{N}\hat{H}^N_{\rm ecqf}\hat{P}_{N}+
  \hat{P}^{\dag}_{N+2}\left(\hat{H}^{N+2}_{\rm ecqf}+
    \Delta^{N+2}\right)\hat{P}_{N+2}\
  +\hat{V}_{\rm mix}^{N,N+2}~,
\label{eq:ibmhamiltonian}
\end{equation}
where $\hat{P}_{N}$ and $\hat{P}_{N+2}$ are projection operators onto
the $[N]$ and the $[N+2]$ boson spaces, 
respectively, $\hat{V}_{\rm mix}^{N,N+2}$  describes
the mixing between the $[N]$ and the $[N+2]$ boson subspaces, and
\begin{equation}
  \hat{H}^i_{\rm ecqf}=\varepsilon_i \hat{n}_d+\kappa'_i
  \hat{L}\cdot\hat{L}+
  \kappa_i
  \hat{Q}(\chi_i)\cdot\hat{Q}(\chi_i), \label{eq:cqfhamiltonian}
\end{equation}
is the extended consistent-Q Hamiltonian (ECQF) \cite{warner83} with $i=N,N+2$,
$\hat{n}_d$ the $d$ boson number operator, 
\begin{equation}
  \hat{L}_\mu=[d^\dag\times\tilde{d}]^{(1)}_\mu ,
\label{eq:loperator}
\end{equation}
the angular momentum operator, and
\begin{equation}
  \hat{Q}_\mu(\chi_i)=[s^\dag\times\tilde{d}+ d^\dag\times
  s]^{(2)}_\mu+\chi_i[d^\dag\times\tilde{d}]^{(2)}_\mu~,\label{eq:quadrupoleop}
\end{equation}
the quadrupole operator. 
We are not considering the most general
IBM Hamiltonian in each Hilbert space, [N] and [N+2], but we are
restricting ourselves to an ECQF formalism \cite{warner83,lipas85} in each subspace. This
approach has been shown to be a rather good approximation in many calculations.

The parameter $\Delta^{N+2}$ can be
associated with the energy needed to excite two particles across the
$Z=82$ shell gap, corrected for the pairing interaction gain and including
monopole effects \cite{hey87,hey85}.
The operator $\hat{V}_{\rm mix}^{N,N+2}$ describes the mixing between
the $N$ and the $N+2$ configurations and is defined as
\begin{equation}
  \hat{V}_{\rm mix}^{N,N+2}=w_0^{N,N+2}(s^\dag\times s^\dag + s\times
  s)+w_2^{N,N+2} (d^\dag\times d^\dag+\tilde{d}\times \tilde{d})^{(0)}.
\label{eq:vmix}
\end{equation}

The wave function in the IBM-CM can then be described as follows
\begin{eqnarray}
\Psi(k,JM) &=& \sum_{i} a^{k}_i(J;N) \psi((sd)^{N}_{i};JM) 
\nonumber\\
&+& \
\sum_{j} b^{k}_j(J;N+2)\psi((sd)^{N+2}_{j};JM)~.
\end{eqnarray}

The $E2$ transition operator for two-configuration mixing is
subsequently defined as
\begin{equation}
  \hat{T}(E2)_\mu=\sum_{i=N,N+2} e_i
  \hat{P}_i^\dag\hat{Q}_\mu(\chi_i)\hat{P}_i~,\label{eq:e2operator}
\end{equation}
where the $e_i$ ($i=N,N+2$) are the effective boson charges and
$\hat{Q}_\mu(\chi_i)$ is the quadrupole operator defined in Eq.~(\ref{eq:quadrupoleop}).

The starting point of the present IBM-CM mixing analysis is the work 
by Harder {\it et al.}~\cite{harder97}, 
in which a schematic IBM mixing calculation with a fixed Hamiltonian
along the whole chain of the Pt nuclei  was carried out. The parameters used in that
calculation are presented in Table \ref{tab-par-piet}.
\begin{table}
\begin{center}
\begin{tabular}{|c|c|c|c|c|c|}
\hline
      &$\varepsilon_i$&$\kappa'_i$&$\kappa_i$&$\chi_{i}$&$e_{i}$\\
\hline
$N$   &540            &0          &-27       &0.25     & 0.16  \\
$N+2$ &0              &10         &-22       &-0.45    &0.13   \\      
\hline
\end{tabular}
\end{center}
\caption{Hamiltonian and $\hat{T}(E2)$ parameters from Harder {\it et al.}
  \cite{harder97}. 
All the quantities are given
in keV, 
except for $\chi_{i}$ which are dimensionless and for
the effective charges, $e_{i}$, which are
given in 
e$\cdot$b. The remaining parameters of the Hamiltonian are: $\Delta^{N+2}=1400$ keV and 
$w_0^{N,N+2}=w_2^{N,N+2}=50$ keV.}
\label{tab-par-piet}
\end{table}

Although these calculations were able to reproduce the evolution of the structure 
of the low-lying states in a qualitative way, they do not allow for a
reproduction of the finer details. Besides, $B(E2)$ values have not been calculated in
that study. In Fig.~\ref{fig-piet-mix}, we represent
the major results from \cite{harder97}, in particular 
the excitation energies,
first without the mixing term and relative to the first regular $0^+$
state (upper panel) in which the behavior of the regular and intruder
states stands out clearly, and secondly, including a mixing term equal to
$50$ keV (lower panel). In the lower panel of the figure, the energies are given relative to 
the $0^+$ state with the highest percentage of the regular N-boson subspace. 
In Fig.~\ref{fig-piet-mix-wf} we represent
that part of the wave function contained within the N-boson subspace, defined as
the sum of the squared amplitudes, or weight $w^k(J,N) \equiv \sum_{i}\mid a^{k}_i(J;N)\mid ^2$, of
the two lowest-lying states $(k=1,2)$ for the most important angular momentum values 
that show up in the low-energy spectrum
corresponding to the lower panel of Fig.~\ref{fig-piet-mix}.

Several features are highlighted by inspection of 
Figs.~\ref{fig-piet-mix} and \ref{fig-piet-mix-wf}. First of
all, one observes a rapid change in the structure of the states, with
the lowest states resulting as mainly regular at the beginning (end) of the shell,
while the intruder character becomes dominant at the mid-shell, $A=182~(N=104)$ region. 
The first state for each $L$ is mainly regular at the beginning (end)
of the shell, while mainly intruder at the mid-shell. For the second
state the situation is rather the opposite. It
is mainly an intruder state at the beginning (end) of the shell, while regular
at the mid-shell region. In the case of the $5^+$ and $7^+$ states, they
behave over the whole mass-range as intruder states, with a minimal content of the regular
component at the mid-shell. 
Secondly, the energy spectra are symmetric
with respect to the mid-shell point at $N=104$ with an unrealistic kink appearing 
at the mid-shell position, while the real data do not exhibit such a pronounced
behavior. As a conclusion, it turns out that one would need a 
strong mixing term in order to approximately reproduce
the experimental data. Finally, it becomes clear that the parameters from \cite{harder97}
should be fine tuned in order to reproduce the experimental data
(excitation energy and 
$B(E2)$ values) quantitatively.

An alternative method to analyze this mass region was proposed
recently by
McCutchan {\it et al.}~\cite{cutcham05}, in which the Pt nuclei were treated
as consisting of just four proton holes and a number of valence
neutrons, 
moving outside of the $Z=82$, $N=82$ doubly-closed shell.
Thus, proton excitations across the $Z=82$ closed shell were not taken into account
explicitly. In this single-space approach, considering $N$ bosons,
McCutchan {\it et al.}~used the ECQF
\cite{warner83,lipas85} with the Hamiltonian \cite{werner02}
\begin{equation}
  \label{Qhamiltoniantriangle}
  \hat{H}(\zeta)=c((1-\zeta)\hat{n}_d - \frac{\zeta}{4N} \hat{Q}(\chi).\hat{Q}(\chi))~ ,
\end{equation}
\noindent
where the quadrupole operator is given by
Eq.~(\ref{eq:quadrupoleop}). 
The parameter $c$
is a general energy scaling factor, $N$ is the number of s and d bosons and
$\zeta$ and $\chi$ are two structural parameters, describing the
spherical-deformed transition and the prolate-oblate transition,
respectively. Note that this Hamiltonian can also be rewritten using the
parameters $\varepsilon_d$ and $\kappa$ 
(see equation (\ref{eq:cqfhamiltonian})). 

In the present case, one considers a limited number of basis states, 
{\it i.e.,}~only the components with $N$ bosons. 
This basis spans the smaller model space and the corresponding  model wave function can be expressed as
\begin{equation}
\Psi'(k,JM) = \sum_{l} a'^{k}_{l}(J;N) \psi((sd)^{N}_{l};JM) ~.
\end{equation}

In section \ref{sec-fit-procedure} we present the methods used in order to 
determine the parameters
appearing in the IBM-CM Hamiltonian as well as in the $\hat{T}(E2)$ operator.
We discuss the resulting energy spectra and   
the $B(E2)$ reduced transition probabilities, and carry out a detailed comparison
with both, the experimental data and with the ECQF calculations \cite{cutcham05}.

\subsection{The fitting procedure: energy spectra and absolute $B(E2)$ reduced 
transition probabilities}
\label{sec-fit-procedure}

Here, we present the way in which the parameters of the Hamiltonian
(\ref{eq:ibmhamiltonian}), (\ref{eq:cqfhamiltonian}), and (\ref{eq:vmix}) 
and the effective charges 
in the $\hat{T}(E2)$ transition operator (\ref{eq:e2operator}) have been determined.

In order to compare with the calculations carried out by McCutchan {\it et al.}~\cite{cutcham05},
who studied besides the yrast levels, a number of non-yrast levels and
the corresponding $B(E2)$ values, we have to
carry out a more detailed calculation within the IBM-CM approach, going beyond the more
schematic study carried out by  Harder {\it et al.}~\cite{harder97}.

We concentrate on the range $^{172}$Pt to $^{194}$Pt thereby covering
a major part of the neutron $N=82-126$ shell. This interval also corresponds to
the same set of isotopes analyzed in references 
\cite{cutcham05} and \cite{harder97}.

In the fitting procedure carried out here, we try to obtain the best possible agreement
with the experimental data including both the excitation energies
and the $B(E2)$ reduced transition probabilities. 
Using the expression of the IBM-CM Hamiltonian, as given in equation (\ref{eq:ibmhamiltonian}),
and of the $E2$ operator, as given in Eq.~(\ref{eq:e2operator}), in the most general case 
thirteen parameters show up.  
We impose a constraint of using parameters that change smoothly  
in passing from isotope to isotope. For the regular Hamiltonian, we
have constrained one of the parameters, {\it i.e.}, $\chi_N=0$, while for the intruder 
Hamiltonian we have fixed the relative d-boson energy to the value
$\varepsilon_{N+2}=0$. 
These constraints follow from a number of test calculations that were carried out
in which no substantial improvement in the value of $\chi^2$ (see
Eq.~(\ref{chi2})) was obtained if we 
allowed $\varepsilon_{N+2}\neq 0$ or $\chi_N \neq 0$. Note that the
constraint $\varepsilon_{N+2}=0$ is also supported by \cite{harder97}. 
On the other hand, we have kept the value that describes the
energy needed to create an extra particle-hole pair ($2$ extra bosons) constant, 
{\it i.e.}, $\Delta^{N+2}=1400$ keV, and have also put the constraint of keeping the
mixing strengths constant, {\it i.e.}, $w_0^{N,N+2}=w_2^{N,N+2}=50$ keV for all the Pt isotopes. 
Those parameter values have been shown to be quite appropriate in this part
of the nuclear mass region \cite{harder97,king98}, although the choice of 
the mixing strength remains somewhat arbitrary \cite{harder97}. 
We also have to determine for each isotope the effective charges of the $E2$ operator.
This finally leads to eight parameters to be varied in each nucleus.

The $\chi^2$ test is used in the fitting procedure in order to extract the optimal solution. 
The $\chi^2$ function is defined in the standard way as
\begin{equation}
  \label{chi2}
  \chi^2=\frac{1}{N_{data}-N_{par}}\sum_{i=1}^{N_{data}}\frac{(X_i
    (data)-X_i (IBM))^2}{\sigma_i^2},
\end{equation} 
where $N_{data}$ is the number of experimental data,
$N_{par}$ is the number of parameters used in the IBM fit, $X_i(data)$
describes the experimental excitation energy of a given experimental energy level (or an experimental 
$B(E2)$ value), $X_i(IBM)$ denotes the corresponding calculated IBM-CM value,
and $\sigma_i$ is an error assigned to each $X_i(data)$ point. 

The $\chi^2$ function is defined as a sum over all data points including excitation
energies as well as absolute $B(E2)$ values. The minimization is carried out using 
$\varepsilon_N$, $\kappa'_N$,
$\kappa_N$, $\kappa'_{N+2}$, $\kappa_{N+2}$, $\chi_{N+2}$, $e_{N}$ and
$e_{N+2}$ as free parameters, having fixed $\chi_{N}=0$,
$\varepsilon_{N+2}=0$, $\Delta^{N+2}=1400$
keV and $w_0^{N,N+2}=w_2^{N,N+2}=50$ keV as described before. We minimize the $\chi^2$
function for each isotope separately using the package MINUIT \cite{minuit} which allows to
minimize any multi-variable function.
In some of the lighter Pt isotopes, due to the small number of experimental data, the
values of some of the free 
parameters could not be fixed unambiguously using the above fitting procedure.

As input values, we have used the excitation energies of the levels presented in 
Table \ref{tab-energ-fit}. In this table we also give the corresponding $\sigma$ values. 
We stress
that the $\sigma$ values do not correspond
to experimental error bars, but they are related with the expected
accuracy of the IBM-CM calculation to reproduce a particular experimental data point. 
Thus, they act as a guide so that a given calculated level converges towards the
corresponding experimental level.
The $\sigma$ ($0.1$ keV) value for the $2_1^+$ state guarantees
the exact reproduction of this experimental most important excitation energy, {\it
  i.e.,}~the whole energy spectrum is normalized to this experimental energy. As in
reference \cite{cutcham05}, the states $4_1^+, 0_2^+$ and $2_2^+$ are
considered as the most important ones to be reproduced ($\sigma=1$
keV). The group of states  
$4_2^+, 4_3^+, 6_1^+, 8_1^+$ and  $3_1^+$ ($\sigma=10$
keV) and $2_3^+$ ($\sigma=100$ keV) should also be well
reproduced by the calculation to guarantee a correct moment of
inertia for the yrast band and the structure of the pseudo-$\gamma$
and $0_2^+$ bands. 
In the case of the $E2$ transition rates, we have used the available
experimental data involving the states presented in Table \ref{tab-energ-fit},
restricted to those E2 transitions for which absolute $B(E2)$ values are known.
Additionally we have taken a value of $\sigma$ corresponding to $10\%$ of these 
$B(E2)$ values. 
In view of the large number of relative $B(E2)$ values, we have derived optimal effective charges
using the same fitting scheme as before, but now keeping the parameters in the IBM-CM Hamiltonian fixed.
Here, we include the relative $B(E2)$ values and for these data 
we have used a relative error of a 20\%.  
The experimental data that we have used are taken from the Nuclear Data Sheets (NDS), Adopted Values references
\cite{Sing95,Brow99,Basu06,Brow94,Wu03,Sing95b,
Fire89,Bagl03,Sing02,Sing03,Bagl98,Sing06}, unless the latest issue was published more than 10 years ago.
Therefore, for the mass number $A=182, 184$ and $192$, the experimental data were taken from 
\cite{davidson99,baglin09,cutcham08}, respectively.

\begin{table}
\begin{center}
  \begin{tabular}{|c|c|}
    \hline
    Precision (keV) & States  \\
    \hline
    $\sigma=0.1$& $2_1^+$ \\
    $\sigma=1$ & $4_1^+, 0_2^+, 2_2^+$\\
    $\sigma=10$  & $4_2^+, 4_3^+, 6_1^+, 8_1^+, 3_1^+$ \\
    $\sigma=100$   & $2_3^+$ \\
    \hline
  \end{tabular}
\end{center}
  \caption{Energy levels, characterized by $J^{\pi}_i$, included in
    the energy fit, if known, and the assigned $\sigma$ values in keV.}
  \label{tab-energ-fit}
\end{table}

The fitting procedure, outlined before, has resulted in the values of the parameters for the IBM-CM Hamiltonian,
as given in Table \ref{tab-fit-par-mix}. 
Note that some of the Hamiltonian parameters,
especially for $^{172}$Pt and $^{174}$Pt, remain arbitrary due to
the lack of experimental data. 
In the case of $^{172}$Pt and  $^{174}$Pt
the value of the effective charges cannot be determined
because not a single $B(E2)$ value is known. 
In the case of $^{182}$Pt
the absolute value of the effective charges cannot be determined
because any absolute $B(E2)$ value is known. Therefore, the given
values are dimensionless.
\begin{table}
\begin{center}
\begin{tabular}{|c||c|c|c||c|c|c||c|c|}
\hline
Nucleus&$\varepsilon_N$&$\kappa'_N$&$\kappa_N$&$\kappa'_{N+2}$&
$\kappa_{N+2}$&$\chi_{N+2}$&$e_{N}$&$e_{N+2}$\\
\hline
$^{172}$Pt&725.0&0.00  &-39.47&0.00 &-22.87&-0.38 &-&-\\
$^{174}$Pt&701.2&0.00  &-31.60&0.00 &-21.82&-0.30 &-&-\\  
$^{176}$Pt&683.4&1.04  &-37.62&5.24 &-23.56&-0.75 &1.86&1.63\\  
$^{178}$Pt&753.8&-2.31 &-37.45&5.27 &-25.17&-0.55 &3.21&1.52\\  
$^{180}$Pt&999.3&-15.14&-37.34&6.57 &-25.14&-0.32 &1.29 &1.94 \\  
$^{182}$Pt&939.9&-6.70 &-35.39&7.03 &-23.50&-0.31 &1&1.1\\  
$^{184}$Pt&750.6&1.47  &-32.66&6.64 &-23.89&-0.34 &1.14 &1.71 \\  
$^{186}$Pt&675.3&3.17  &-30.50&7.29 &-24.23&-0.32 &1.44 &1.67 \\  
$^{188}$Pt&483.2&4.94  &-37.38&6.67 &-31.47&-0.11 &1.66 &1.66 \\  
$^{190}$Pt&338.7&19.33 &-34.62&0.83 &-32.51&0.00  &1.50 &1.50 \\  
$^{192}$Pt&314.9&12.01 &-45.32&-8.82&-38.84&0.00  &1.68&1.77\\  
$^{194}$Pt&370.9&6.67  &-38.26&6.52 &-31.02&0.00  &1.97&0.25\\  
\hline
\end{tabular}
\end{center}
\caption{Hamiltonian and $\hat{T}(E2)$ parameters resulting from the present study.
 All quantities have the dimension of energy (given in units of keV),
except $\chi_{N+2}$ which is dimensionless and $e_{N}$ and $e_{N+2}$
which are given in units $\sqrt{\mbox{W.u.}}$, except for $^{182}$Pt which
are dimensionless (see text).  
The remaining parameters of the
Hamiltonian, {\it i.e.,} $\chi_N$ and $\varepsilon_{N+2}$} are equal to
zero, except  $\Delta^{N+2}=1400$ keV and $w_0^{N,N+2}=w_2^{N,N+2}=50$ keV.
\label{tab-fit-par-mix}
\end{table}

In the ECQF fit carried out  by McCutchan {\it et al.}~\cite{cutcham05}, the
parameters they obtained are given in Table \ref{tab-fit-par-cqf}
\cite{Mccu-pri}. We present the values of $\zeta$ and $\chi$ and also 
the corresponding values of $\varepsilon$ and
$\kappa$ obtained after normalizing $E(2_1^+)$ to the experimental value. 
McCutchan {\it et al.}~used the ratios $E(4_1^+)/E(2_1^+)$, $E(2_\gamma^+)/E(2_1^+)$
and $E(0_2^+)/E(2_1^+)$ as well as some $B(E2)$ ratios, if known, such as
$B(E2;2^+_2 \rightarrow 0^+_1)/B(E2;2^+_2 \rightarrow 2^+_1)$ and
$B(E2;2^+_3 \rightarrow 0^+_1)/B(E2;2^+_3 \rightarrow 2^+_1)$,
in order to determine the Hamiltonian parameters as given in equation
(\ref{Qhamiltoniantriangle}). The value of the 
effective charge of the $\hat{T}(E2)$
operator was fixed 
to reproduce $B(E2;2_1^+\rightarrow 0_1^+)$
except for $^{178}$Pt where $B(E2;4_1^+\rightarrow 2_1^+)$ has been used.
The ECQF fit essentially consists of four free parameters to be varied in
each nucleus, separately, which is clearly less than in the IBM-CM calculations.
\begin{table}
\begin{center}
\begin{tabular}{|c||c||c|c|c|c|}
\hline
Nucleus&$\zeta$&$\varepsilon$&$\kappa$&$\chi$&$e$\\
\hline
$^{172}$Pt&0.49&856.9&-25.7&-1.20&-\\
$^{174}$Pt&0.51&811.6&-23.5&-1.10&-\\  
$^{176}$Pt&0.47&480.8&-10.7&-1.10&2.22\\  
$^{178}$Pt&0.55&472.5&-13.1&-1.00&2.09\\  
$^{180}$Pt&0.57&480.9&-13.3&-0.90&2.17\\  
$^{182}$Pt&0.57&516.3&-13.2&-0.87&-\\  
$^{184}$Pt&0.57&480.0&-13.3&-0.84&1.90\\  
$^{186}$Pt&0.59&497.0&-16.3&-0.70&1.88\\  
$^{188}$Pt&0.64&537.2&-23.9&-0.30&1.91\\  
$^{190}$Pt&0.66&526.6&-28.4&-0.20&1.70\\  
$^{192}$Pt&0.72&481.6&-38.7&-0.10&1.82\\  
$^{194}$Pt&0.74&431.3&-43.8&-0.10&1.86\\  
\hline
\end{tabular}
\end{center}
\caption{ECQF Hamiltonian parameters and $\hat{T}(E2)$ effective
  charge \cite{cutcham05,Mccu-pri}. 
  All the quantities have the dimension of energy (units of keV),
  except $\zeta$ and $\chi$ which are dimensionless, and $e$ given in 
  units $\sqrt{\mbox{W.u.}}$ Note that $\zeta$ is not an independent
  variable, but depends on $\varepsilon$ and $\kappa$.}
\label{tab-fit-par-cqf}
\end{table}

\begin{figure}[hbt]
  \centering
  \epsfig{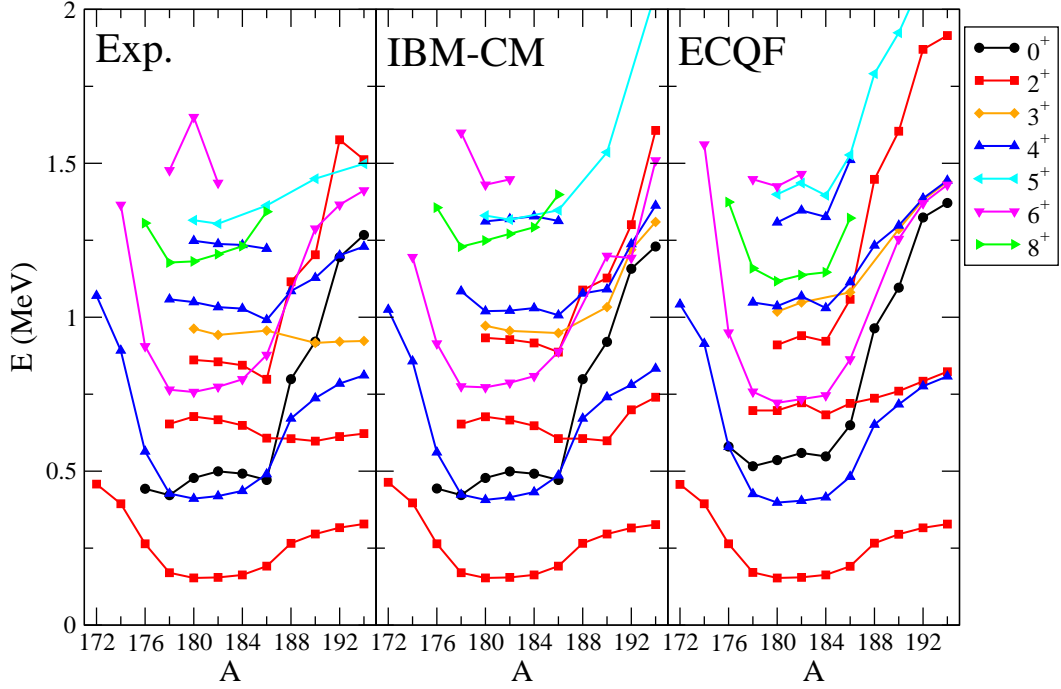} 
  \caption{Experimental excitation energies (up to $E_x \approx 1.5$
    MeV) and theoretical results obtained from the IBM-CM and ECQF
    calculations \cite{cutcham05}.}
  \label{fig-energ-comp}
\end{figure}

\begin{figure}[hbt]
  \centering
  \epsfig{file=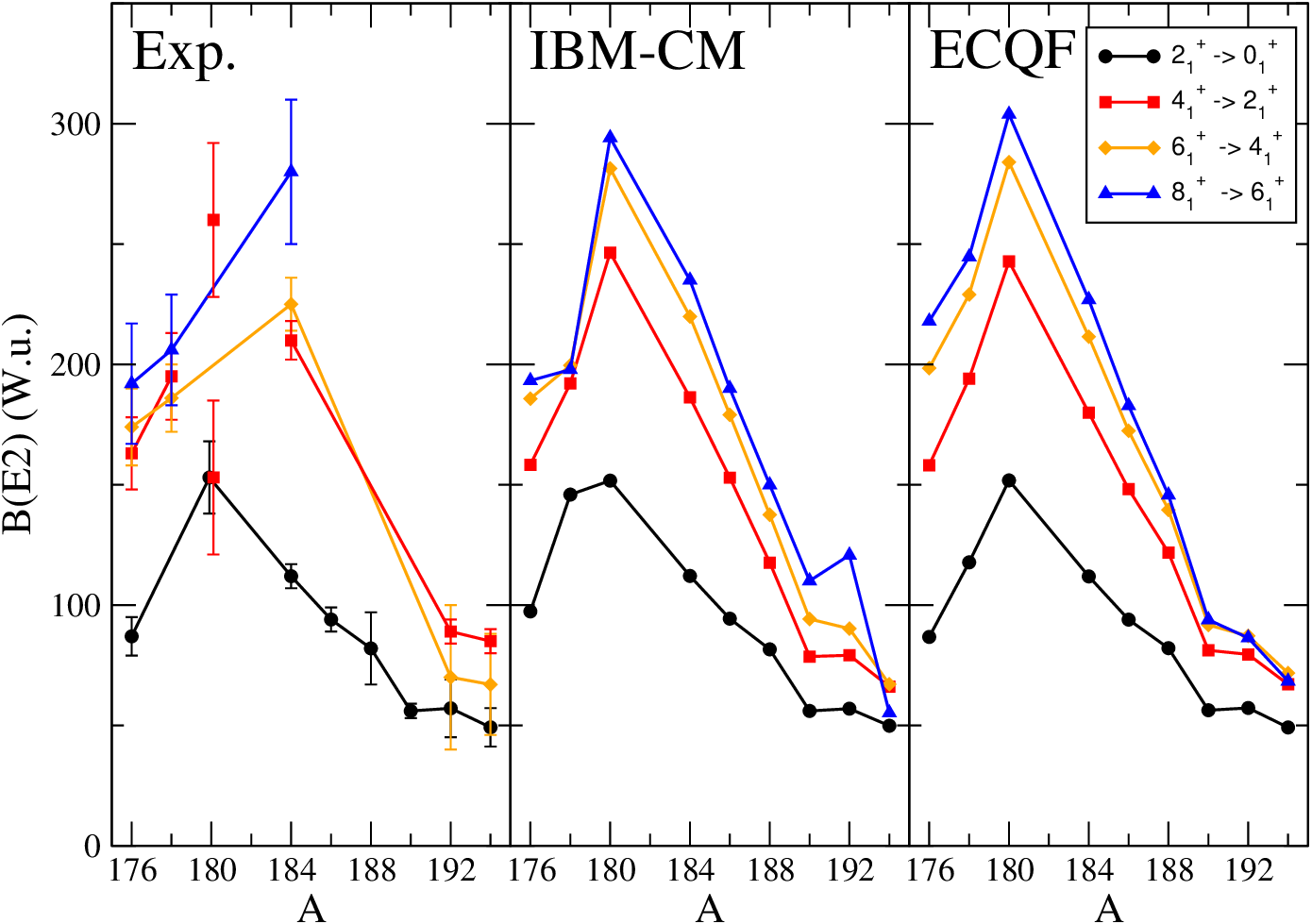,width=14cm} 
  \caption{Comparison of a set of absolute $B(E2)$ reduced transition
    probabilities in the ground band, given in W.u. The left panel corresponds to known
    experimental data, the central panel to the theoretical IBM-CM results
    and the right panel to the theoretical ECQF results.}
  \label{fig-be2-1}
\end{figure}

\begin{figure}[hbt]
  \centering
  \epsfig{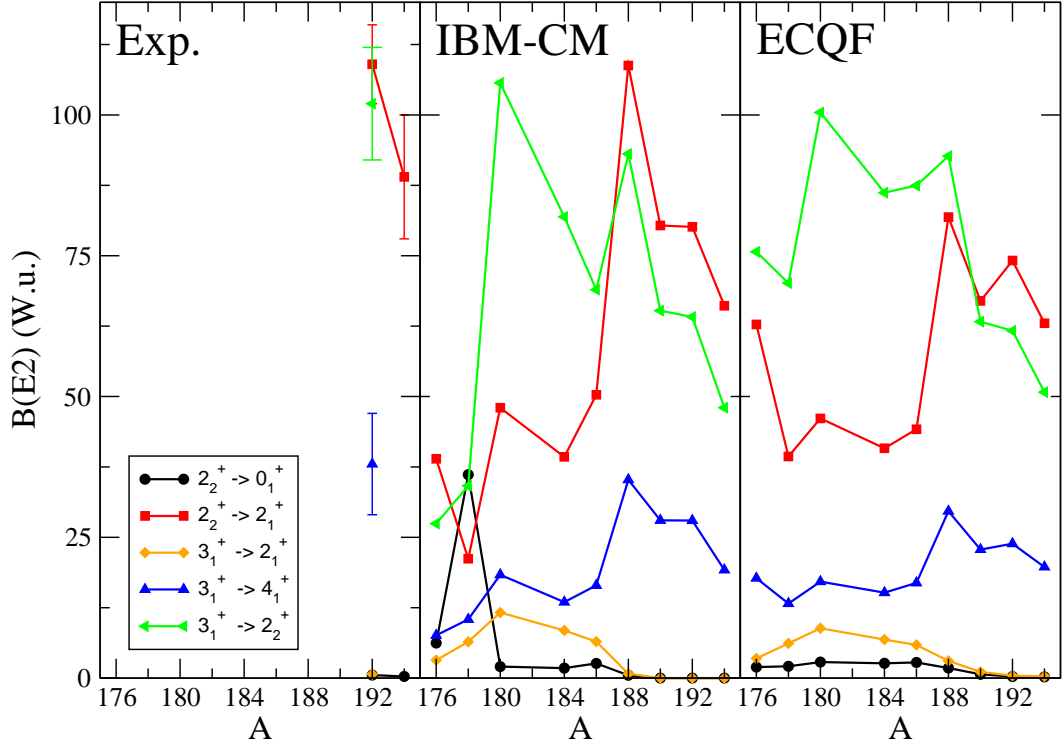} 
  \caption{Comparison of a set of absolute $B(E2)$ reduced transition
    probabilities, given in W.u. The left panel corresponds to 
    experimental data, the central panel to the IBM-CM results
    and the right panel to the ECQF results.}
  \label{fig-be2-2}
\end{figure}

\begin{figure}[hbt]
  \centering
  \epsfig{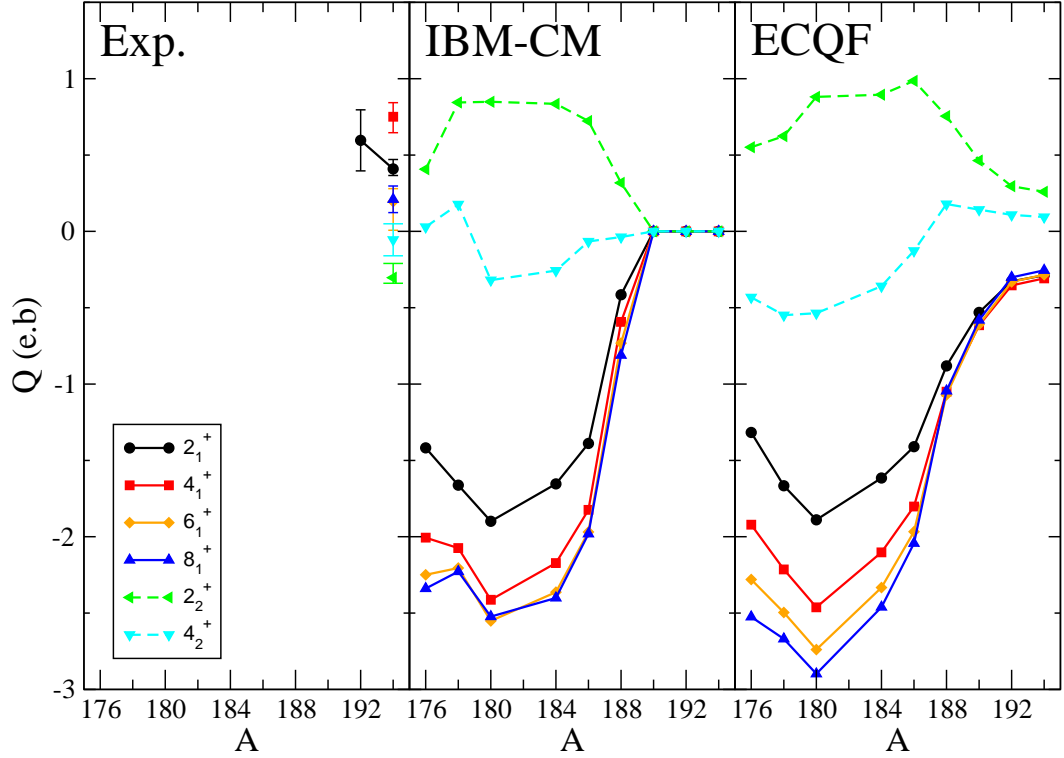} 
  \caption{Comparison of the quadrupole moments for the ground band and the
   $2^+_2,4^+_2$ states, given in units 
    e$\cdot$b.  The left panel corresponds to 
    experimental data, the central panel to the IBM-CM results
    and the right panel to the ECQF results.}
  \label{fig-be2-3}
\end{figure}

\begin{figure}[hbt]
  \centering
  \epsfig{file=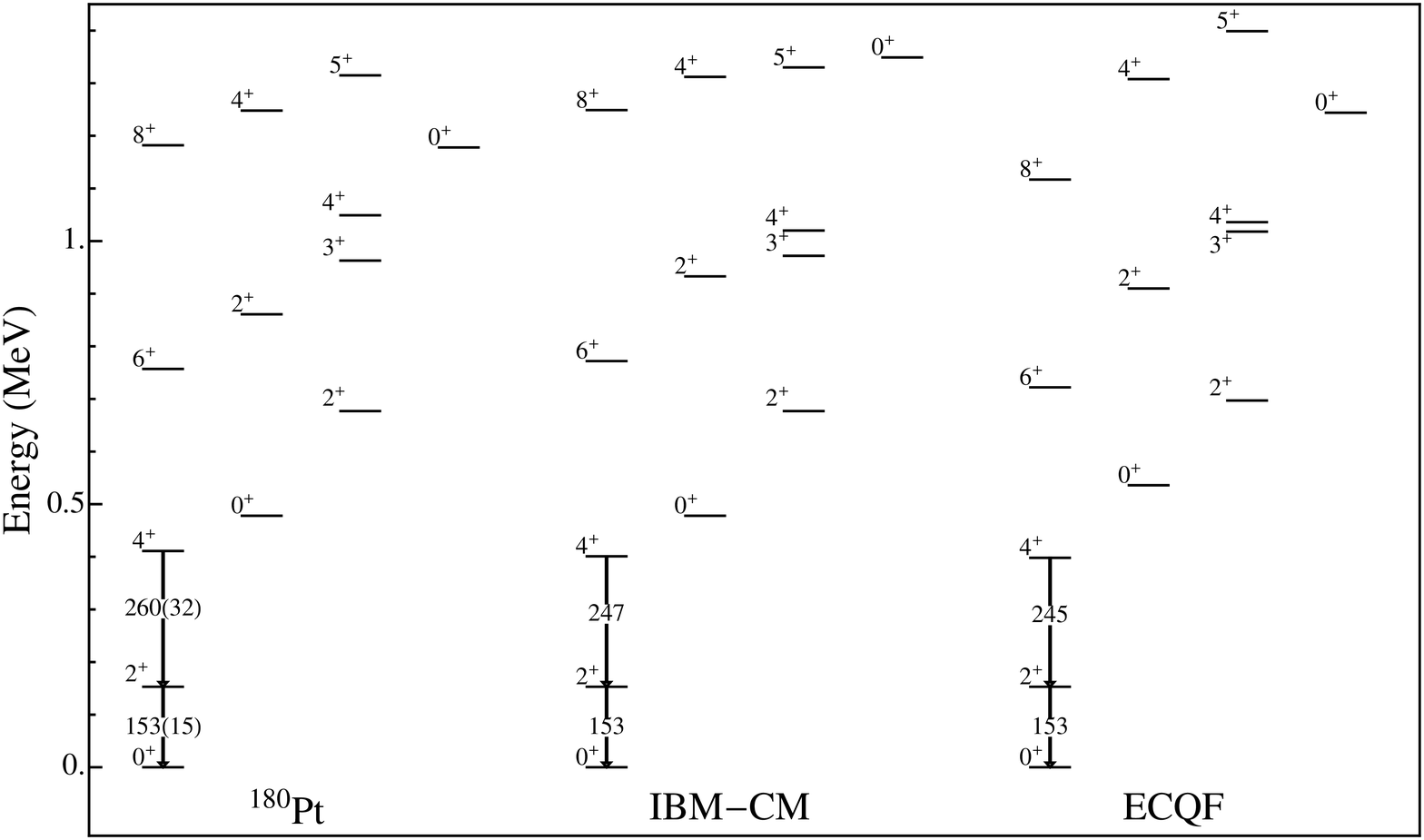,width=14cm} 
  \caption{Detailed comparison, including excitation energies and
    absolute $B(E2)$ values, if known, between experimental data, IBM-CM
    and ECQF for $^{180}$Pt.} 
  \label{fig-180pt}
\end{figure}

\subsection{Comparing the IBM-CM and ECQF: energy spectra and electric quadrupole  
properties}
\label{sec-fit-compare}

In the present subsection, we compare the energy spectra as obtained from the IBM-CM 
and from the ECQF with the 
experimental energy spectra, for the limited data set (with excitation energy $E_x$ less than
$\approx$ 1.5 MeV), so as to be able to carry out a detailed comparison between
both model calculations, as well as between each of the calculations and
the experimental data (see Fig.~\ref{fig-energ-comp}). It becomes clear, 
by inspecting these results, that both approaches 
seem equally good in describing the experimental energies up to an
excitation energy $E_x \approx 1.5$ MeV, although, in general, the IBM-CM calculation
provides certain improvements with respect to the ECQF results. 
A striking result is that the experimental energy spectra in $^{178-186}$Pt are virtually identical,
pointing towards a common underlying collective structure.

A more detailed test of the content of the resulting wave functions
comes from the $B(E2)$ reduced transition probabilities and from the
electric quadrupole moments. 
The trends for a number of important E2 transitions and for electric quadrupole
moments are presented in Figs.~\ref{fig-be2-1}, \ref{fig-be2-2}, 
and \ref{fig-be2-3}.  
In Fig.~\ref{fig-be2-1}, we show the absolute $B(E2)$ values
for the yrast band. First of all, one observes close similarities between
both theoretical approaches and good agreement with the experimental
data (we show the two non-overlapping $B(E2;4^+_1 \rightarrow 2^+_1)$
experimental values for $^{180}$Pt). 
In particular, the increase in the $B(E2)$
values with increasing angular momentum $J$, as well as a saturation near the higher spin $J=8$ 
value is well reproduced. 
Moreover, when moving from the lighter isotopes up to the mid-shell region a steady increase
in the $B(E2)$ values shows up, followed by a rather smooth decrease when moving
towards the heavier Pt isotopes. In Fig.~\ref{fig-be2-2}, a number of
interband $B(E2)$ values are plotted. Here, some clear
differences between IBM-CM and ECQF show up, especially for the   
$B(E2;2^+_2 \rightarrow 0^+_1)$, $B(E2;2^+_2 \rightarrow 2^+_1)$ and 
$B(E2;3^+_1 \rightarrow 2^+_2)$
values in the lighter Pt isotopes. Progressing towards the heavier
isotopes, both approaches provide overall similar results. In comparing
with the scarce available experimental data (only known for the heavier Pt isotopes), 
the agreement is satisfactory. 
In Fig.~\ref{fig-be2-3}, the quadrupole moments
of some yrast and quasi-$\gamma$ band states are given. In general,
both theoretical approaches show the same variation with angular momentum $J$
and mass number $A$. Unfortunately, there exist very few data to test this behavior: 
experimental quadrupole moments are only known for the heavier
$^{192,194}$Pt isotopes \cite{Wu96,Ston05}. 
In the lighter isotopes, a negative value of quadrupole moment for
the yrast band state results, which is increasing and approaches 
an almost vanishing value for the heavier Pt isotopes. For the non-yrast
$2^+_2$,  a positive quadrupole moment results, which is decreasing in approaching the heavier isotopes.
Both theoretical models result again in a similar behavior. For the non-yrast
$4^+_2$ state, the ECQF results in a more smooth variation, changing
sign between $A=186$ and $A=188$, as compared
with the IBM-CM results. Both descriptions tend towards a much reduced quadrupole moment in the heavier
isotopes. 
The main differences between both approaches is that, in
general, the ECQF generates a larger range for the quadrupole moments  
(larger negative quadrupole moments for the yrast band and $4^+_2$ state) 
along the whole chain of isotopes and, the quadrupole
moments for the heavier isotopes become zero for the IBM-CM (for all $J$ values)
while different from zero but having opposite signs with respect to the experimental data in the ECQF. 
This effect is a consequence of the differences in the  values of $\chi$ that were used in the IBM-CM (with
$\chi_N$ and $\chi_{N+2}$=0) and ECQF, respectively.  
\begin{figure}[hbt]
  \centering
  \epsfig{file=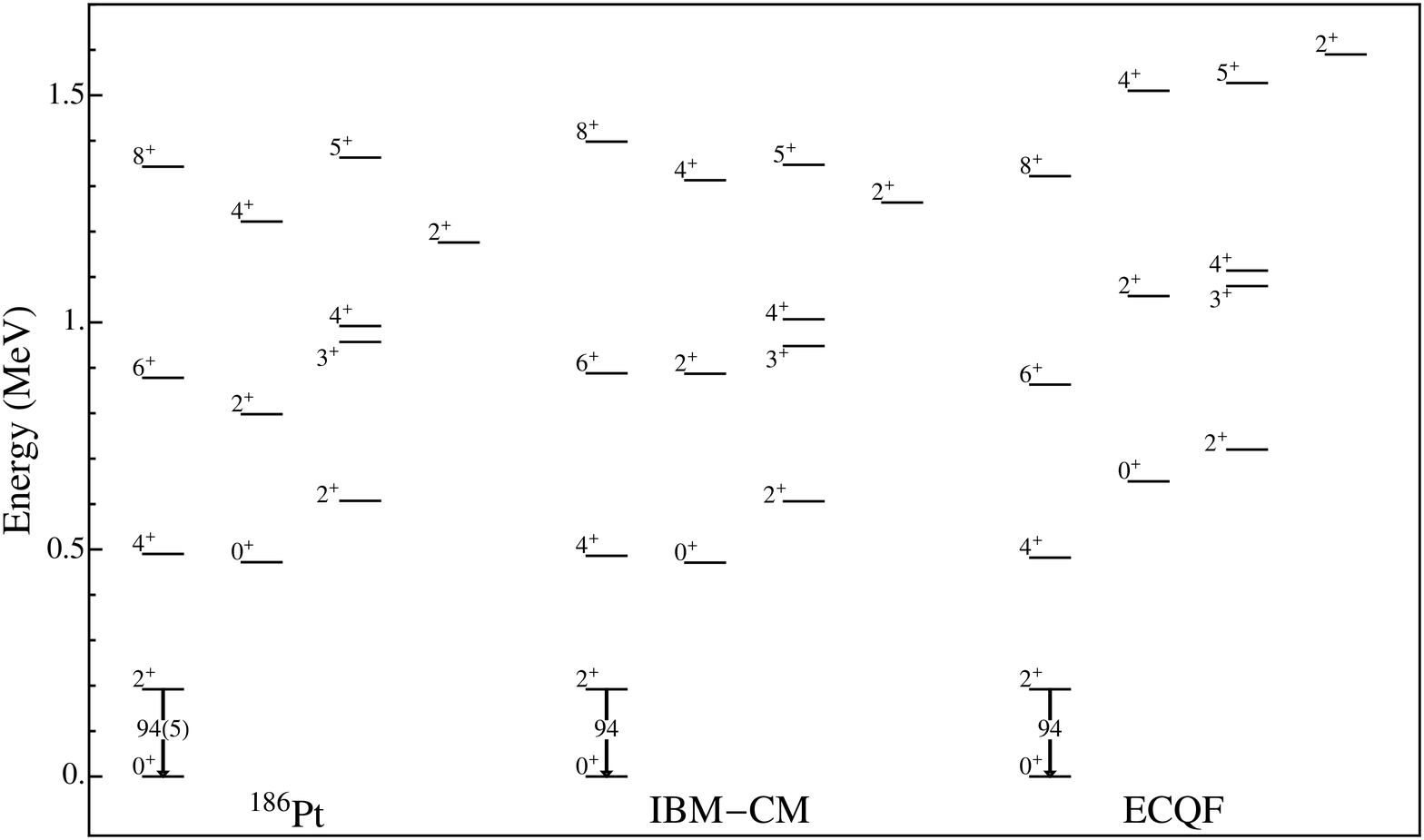,width=14cm} 
  \caption{Detailed comparison, including excitation energies and
    absolute  $B(E2)$ values, if known, between experimental data, IBM-CM
    and ECQF for $^{186}$Pt.}
  \label{fig-186pt}
\end{figure}

In Tables \ref{tab-be2} and \ref{tab-q} and in Tables
\ref{tab-be2-1}, \ref{tab-be2-2}, \ref{tab-be2-3}, and \ref{tab-be2-4}, in appendix \ref{ap-A}, 
we compare the experimental $B(E2)$ values and the electric quadrupole moments 
with the IBM-CM and ECQF theoretical results, respectively. 
For the absolute $B(E2)$ values (Table \ref{tab-be2}) 
we observe an overall similarity between both approaches and a
reasonably good agreement with the experimental data. 
In the case of  $^{178}$Pt, the ECQF $B(E2)$ values along
the yrast band are between $15-20\%$ larger compared to the IBM-CM results and the data.
In $^{180}$Pt, $^{184}$Pt, $^{192}$Pt, and $^{194}$Pt  both theoretical approaches provide
strongly similar values and good agreement with the experimental data. 
However, in $^{194}$Pt, the $E2$ transitions
decaying from the $0^+_2$ state into the $2^+_{1,2}$ states show significant differences
between the ECQF and IBM-CM approach, as well as for the $0^+_4$ to   
$2^+_1$ E2 transition.
This may point towards  a slightly better description of the
structure of the excited $0^+_2$ 
and $0^+_4$ states within the IBM-CM description. However, both the
ECQF and IBM-CM deviate 
strongly from the experimental value for the $0^+_4$ to $2^+_2$ E2 transition.  
Thus, it may also turn out that some of the improvements of the IBM-CM over
the ECQF could be related to the use of two effective charges instead
of one. 
Experiments to extract E0 transition rates decaying from these $0^+$ states are therefore
very important. 
The relative $B(E2)$ values are presented in 
Tables
\ref{tab-be2-1}, \ref{tab-be2-2}, \ref{tab-be2-3}, and \ref{tab-be2-4}
in the appendix \ref{ap-A} in view of the extensive
character of that table. There, we give all data used in order to extract these relative values
starting from the intensities of $\gamma$-transitions, specifying the decay of specific levels.
Regarding the quadrupole moments (Table \ref{tab-q}), where experimental data are
only available for $^{192-194}$Pt, the IBM-CM predicts a vanishing
quadrupole moments for all spin $J$ states. The ECQF predicts finite quadrupole moments but with a
sign opposite to the experimental one. The analysis of the data indicates that the
$^{194}$Pt nucleus corresponds to an oblate but quite $\gamma$-soft structure ~\cite{Wu96}.  
The IBM-CM predicts these nuclei as $\gamma$-unstable nuclei, while the ECQF describes them as prolate nuclei.
These data should allow for an improved description and constrain the parameters in better way. 

\begin{table}
\begin{center}
\begin{tabular}{|c|c|c|c|c|}
\hline
Isotope   &Transition             &Experiment&IBM-CM &ECQF\\
\hline 
$^{176}$Pt&$2_1^+\rightarrow 0_1^+$& 87(8)       &87    & 87 \\  
         &$4_1^+\rightarrow 2_1^+$& 163(15)       &144 & 159  \\   
         &$6_1^+\rightarrow 4_1^+$& 174(16)       &183 & 199   \\   
         &$8_1^+\rightarrow 6_1^+$& 192(25)       &192 & 219   \\   
\hline
$^{178}$Pt&$4_1^+\rightarrow 2_1^+$& 195(18)       &195 &195   \\  
         &$6_1^+\rightarrow 4_1^+$& 186(14)       &199  &230   \\   
         &$8_1^+\rightarrow 6_1^+$& 206(23)       &195  &246  \\   
\hline 
$^{180}$Pt&$2_1^+\rightarrow 0_1^+$& 153(15)       &153  &153  \\  
         &$4_1^+\rightarrow 2_1^+$& 260(32)       &247   &245 \\   
         &                        & 140(30)       &      &    \\
\hline 
$^{184}$Pt&$2_1^+\rightarrow 0_1^+$& 112(5)       &112 & 112   \\  
         &$4_1^+\rightarrow 2_1^+$& 210(8)       &186   &180 \\   
         &$6_1^+\rightarrow 4_1^+$& 225(11)       &220  &212  \\   
         &$8_1^+\rightarrow 6_1^+$& 280(30)       &235  &227  \\   
         &$10_1^+\rightarrow 8_1^+$& 300(50)       &239 &231   \\   
\hline 
$^{186}$Pt&$2_1^+\rightarrow 0_1^+$& 94(5)       &94   &94 \\  
\hline 
$^{188}$Pt&$2_1^+\rightarrow 0_1^+$& 82(15)       &82   &82 \\  
\hline 
$^{190}$Pt&$2_1^+\rightarrow 0_1^+$& 56(3)       &56   &56 \\  
\hline 
$^{192}$Pt&$2_1^+\rightarrow 0_1^+$& 57.1(12)       &57  &57  \\  
         &$2_2^+\rightarrow 0_1^+$& 0.54(4)     &0.0    &0.28\\   
         &$2_2^+\rightarrow 2_1^+$& 109(7)      &80    &74\\   
         &$3_1^+\rightarrow 4_1^+$& 38(9)       &28    &28\\   
         &$3_1^+\rightarrow 2_2^+$& 102(10)     &64   &61 \\   
         &$3_1^+\rightarrow 2_1^+$& 0.68(7)     &0.0   &0.44 \\   
         &$4_1^+\rightarrow 2_1^+$& 89(5)       &79    &79\\   
         &$6_1^+\rightarrow 4_1^+$& 70(30)      &90   &87 \\   
\hline 
$^{194}$Pt&$2_1^+\rightarrow 0_1^+$& 49.2(8)     &49.6 &49  \\  
         &$2_2^+\rightarrow 0_1^+$& 0.29(4)     &0.0  & 0.21  \\   
         &$2_2^+\rightarrow 2_1^+$& 89(11)       &66  &63  \\   
         &$4_1^+\rightarrow 2_1^+$& 85(5)       &66   &67 \\   
         &$4_2^+\rightarrow 4_1^+$& 14       &32      &33\\   
         &$4_2^+\rightarrow 2_1^+$& 0.36(7)     &0.0  &0.0  \\   
         &$4_2^+\rightarrow 2_2^+$& 21(4)       &35   &37 \\   
         &$6_1^+\rightarrow 4_1^+$& 67(21)       &67  &72  \\   
         &$0_2^+\rightarrow 2_1^+$& 0.63(14)     &0.91 &4.5   \\   
         &$0_2^+\rightarrow 2_2^+$& 8.4(19)      &9.2  &39  \\   
         &$0_4^+\rightarrow 2_1^+$& 14.1(12)     &14   &0.02 \\   
         &$0_4^+\rightarrow 2_2^+$& 14.3(14)     &0.0  &0.0  \\   
\hline
\end{tabular}
\end{center}
  \caption{Comparisons of the experimental absolute $B(E2)$ values (given in
    units of W.u.) with
    the IBM-CM Hamiltonian and the ECQF results \cite{cutcham05}.
    Data are taken from  the Nuclear Data Sheets~\cite{Sing95,Brow99,Basu06,Brow94,Wu03,Sing95b,
      Fire89,Bagl03,Sing02,Sing03,Bagl98,Sing06}, complemented with references 
      presented in section \ref{sec-expe}.} 
  \label{tab-be2}
\end{table}

\begin{table}
\begin{center}
\begin{tabular}{|c|c|c|c|c|}
\hline
Isotope   &State             &Experiment&IBM-CM &ECQF\\
\hline 
$^{192}$Pt&$2_1^+$& 0.6(2)$^{a}$        &0    & -0.327 \\  
\hline 
$^{194}$Pt&$2_1^+$& 0.409$(^{+62}_{-43})^{b}$       &0    & -0.288 \\  
          &$4_1^+$& 0.751$(^{+92}_{-105})^{b}$       &0 & -0.308  \\   
          &$6_1^+$& 0.195$(^{+85}_{-188})^{b}$       &0 & -0.284   \\   
          &$8_1^+$& [-0.06,0.28]$^{b}$       &0 & -0.26   \\   
          &$2_2^+$& -0.303$(^{+93}_{-37})^{b}$       &0 & 0.259   \\   
          &$4_2^+$& -0.06(11)$^{b}$       & 0 & 0.09  \\   
\hline
\end{tabular}
\end{center}
  \caption{Comparison of the experimental quadrupole moments (given in
    e$\cdot$b.) with
    the IBM-CM Hamiltonian and the ECQF results \cite{cutcham05}.
    Data are taken from \cite{Ston05} (a) and \cite{Wu96} (b).} 
  \label{tab-q}
\end{table}

\begin{figure}[hbt]
  \centering
  \epsfig{file=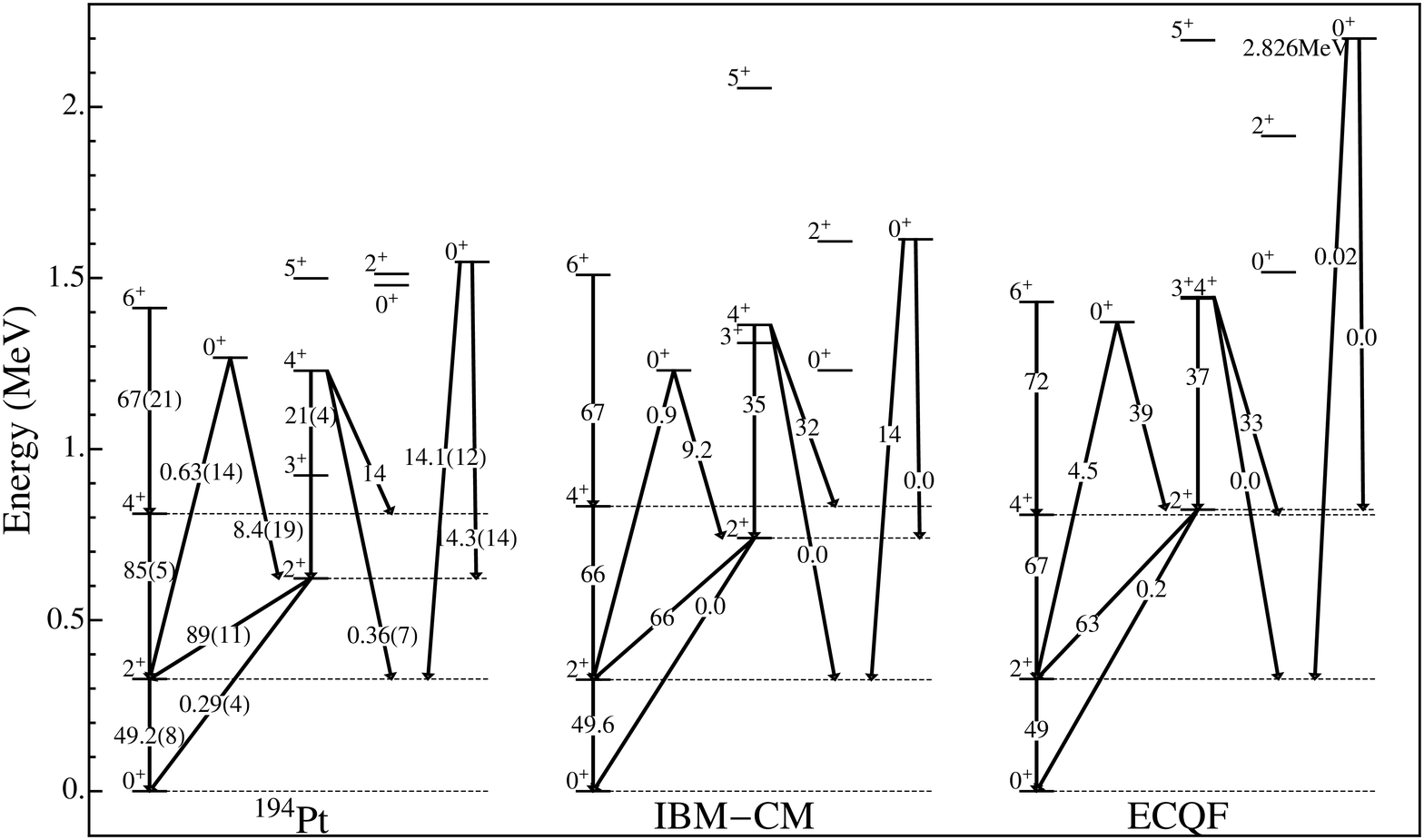,width=14cm} 
  \caption{Detailed comparison, including excitation energies and
    absolute $B(E2)$ values, if known, between experimental data, IBM-CM
    and ECQF for $^{194}$Pt.}
  \label{fig-194pt}
\end{figure}

We now present in Figs.~\ref{fig-180pt}, \ref{fig-186pt}, and
\ref{fig-194pt} the energy spectra (up to $E_x\approx 1.5$ MeV) 
in order to emphasize differences and similarities between the two theoretical
approaches. 

In the case of $^{180}$Pt (Fig.~\ref{fig-180pt}), 
the yrast band  is well described in both calculations up to the $4^+$
level, but slightly high in the IBM-CM while slightly low in the ECQF for the higher-spin members. 
The $0_2^+$ band is correctly described in
both approaches, although the band head is better reproduced by the
IBM-CM. The odd-even staggering is clearly better described by the
IBM-CM.

For $^{186}$Pt (see Fig.~\ref{fig-186pt}), the yrast band is correctly
described in both approaches, IBM-CM and ECQF. The $0_2^+$
band is rather well described by the IBM-CM calculation even though
the spacings in the band are too big compared with the data.
Using the ECQF approximation, the $0_2^+$ band head lies slightly high compared with the
experimental position and here too, the energy spacings are too big compared with the
data.
For the pseudo-$\gamma$ band the situation is
quite similar, with a rather good IBM-CM description, {\it i.e.,} band
head, moment of inertia and odd-even staggering. In the ECQF
calculation, the band head appears somewhat higher in excitation energy while the 
moment of inertia is slightly smaller than the experimental one, even though the 
odd-even staggering is well reproduced.

The comparison for $^{194}$Pt is presented in Fig.~\ref{fig-194pt}.
Here, we observe a good description of the yrast band, including $B(E2)$
values, for both theoretical approaches. The position of the $0_2^+$ state is
correctly reproduced by both calculations. 
The description of the pseudo-$\gamma$
band is only qualitatively reproduced by both approaches: the band
head comes out too high, the even-odd staggering is incorrect and the moment
of inertia is too small. A similar observation results in the description
of the $0_3^+$ band, although, in this case the band-head is better
reproduced by the ECQF. 
 
\subsection{The intruder structure in the IBM-CM results}
\label{sec-fit-intr}

\begin{figure}[hbt]
  \centering
  \epsfig{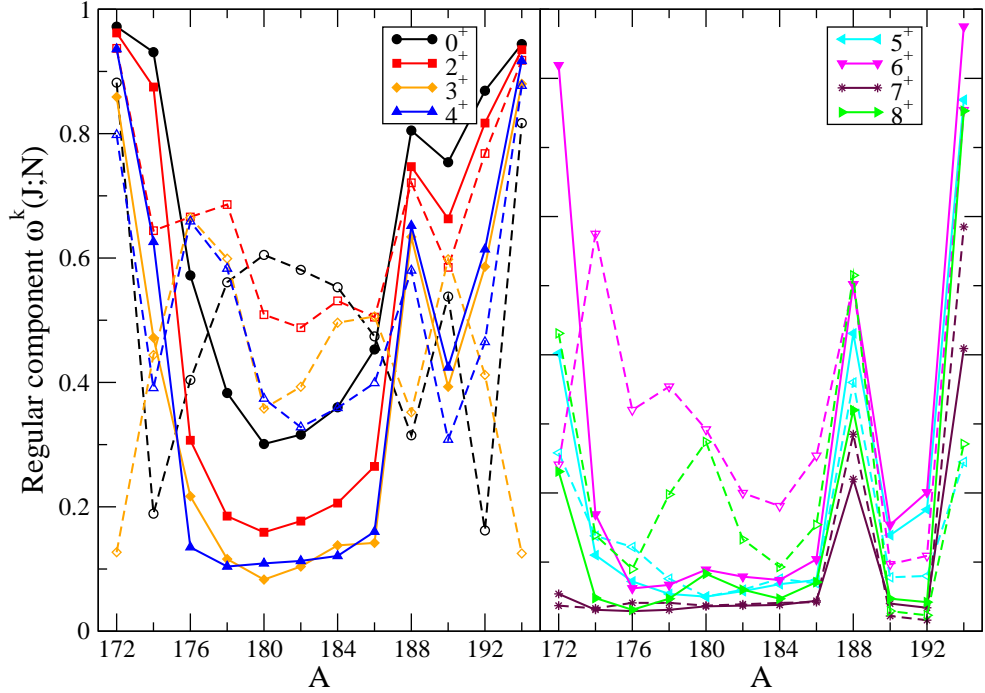} 
  \caption{The regular configuration content, expressed by 
    the weight $w^k(J,N)$ (see text), for the two lowest-lying states
    $(k=1,2)$ for each $J$ value (full lines with
    closed symbols correspond with the first state while dashed lines with open
    symbols correspond with the second state) resulting from the IBM-CM
    calculation, as presented in Fig.~\ref{fig-energ-comp}.}
  \label{fig-wf}
\end{figure}

Having noticed the strong similarities in both excitation energy and
$B(E2)$ reduced transition probabilities considering the particular set 
of levels up to $\approx$ 1.5 MeV,
it is important to study in more detail the particular distribution of the
configurations containing $N$ bosons (and thus also those with $N+2$
bosons) as a function of the changing
mass (neutron) number passing through the  
Pt isotopes and for the various $J^{\pi}$ values
according to the energy spectra as presented in Fig.~\ref{fig-energ-comp}. 
In Fig.~\ref{fig-wf}, we present that part of the wave function
contained within the N-boson subspace, expressed by the
weight $w^k(J,N)$, of the
two lowest-lying  states $(k=1,2)$ for a given angular momentum.
The results obtained for the angular momentum $0^+,2^+,3^+,4^+$ states
(left-hand part) retain the major structure also observed in the calculations carried
out by Harder {\it et al.}~\cite{harder97} and also shown in the 
left-hand part of Fig.~\ref{fig-piet-mix-wf},
although now no longer symmetric with respect to the mid-shell point.
The variation in the weight of the N-boson content also exhibits a far 
more complex structure between neutron (mass) number $N=110$ $(A=188)$ and $N=116$
$(A=194)$. This fact is due to the specific values for the parameters 
characterizing the IBM-CM Hamiltonian for these isotopes. One notices a 
sudden increase in the N-boson weight of the wave functions.
For the higher-spin values $5^+,6^+,7^+,8^+$, the right-hand part of Fig.~\ref{fig-wf} shows serious 
differences when comparing to the right-hand part of Fig.~\ref{fig-piet-mix-wf},
in particular, for the second excited state of each of the spin values shown. Just like
for the left-hand part of the Fig.~\ref{fig-wf}, a rather complex
structure results when passing in between neutron (mass) number $N=110$ $(A=188)$ and
$N=116$ $(A=194)$. 

\begin{figure}[hbt]
  \centering
  \epsfig{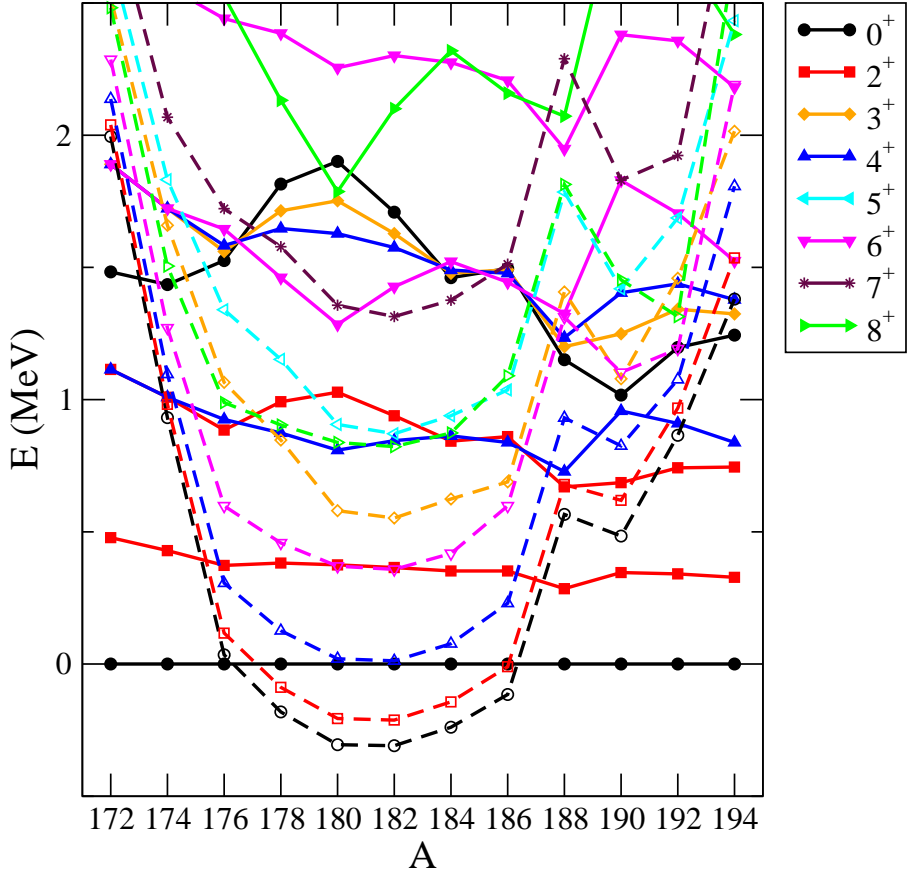} 
  \caption{Energy spectra for the IBM-CM Hamiltonian presented in Table 
    \ref{tab-fit-par-mix}, switching off the mixing term. The two
    lowest-lying regular states and the
    lowest-lying intruder state for each of the angular momenta are shown (full lines with
    closed symbols for the regular states while dashed lines with open
    symbols are used for the intruder states).}
  \label{fig-g5-nomix}
\end{figure}

In order to study more clearly the effects on the energy spectra, induced by the mixing 
term, we recalculate the spectra using the Hamiltonian parameters shown in 
Table \ref{tab-fit-par-mix}, but now 
switching off the mixing term. The spectra are presented in
Fig.~\ref{fig-g5-nomix} 
where we show the two lowest regular and the lowest intruder
state for different angular momentum values. Here, 
we observe a rather flat behavior of
the energy for the regular states. The energy of the intruder
states is smoothly decreasing until the neutron mid-shell value at $N=104$, where it starts
increasing again. 
This effect results mainly from the smooth change of the
Hamiltonian parameters when passing from isotope to isotope.
  
A simultaneous analysis of Figs.~\ref{fig-wf} and
\ref{fig-g5-nomix}, combined with the rules of a simple two-level mixing model, 
allows us to explain the sudden increase
of the regular component for all $J^{\pi}$ values at $A=188$. In 
Fig.~\ref{fig-g5-nomix}, the unperturbed energies, {\it i.e.}, excluding
the mixing term, are plotted. Here, one observes the close
approach of pairs of regular and intruder states with a given
angular momentum, especially in the region around $A=188$. 
The mixing term, coupling the regular ($N$) and intruder 
($N+2$) configurations, can now result in the
interchange of character between the states and therefore in the sudden
increase of the regular component content of the wave function. For states with $J>4$, the
effect is even more dramatic because the unperturbed energy of the intruder configuration
always lies below the unperturbed energy of the regular one and as a
consequence, the interchange in character
with the regular configuration at the point of closest approach is enhanced. Eventually, 
moving towards $A=194$, the unperturbed energies of the intruder configurations are moving up and
cross the energies of the regular configurations. Therefore, as shown
in Fig.~\ref{fig-wf}, from $A=194$  
onwards, the two lowest-lying states for each $J^{\pi}$ value have become regular (N-component,
mainly) states.

\begin{figure}[hbt]
  \centering
  \epsfig{file=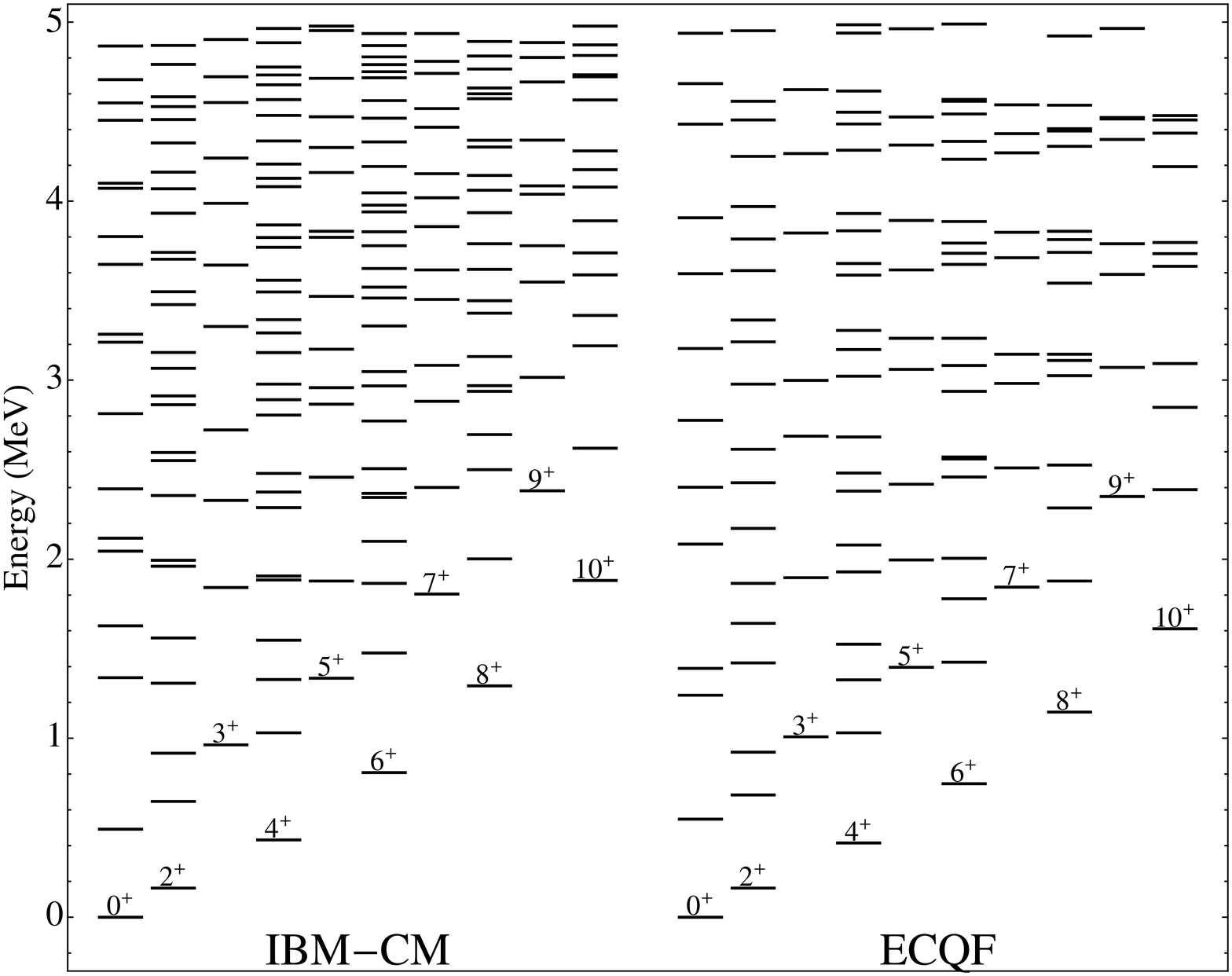,width=14cm} 
  \caption{Comparison of the energy spectra for the IBM-CM and ECQF for $^{184}$Pt up to
    an excitation energy of $E_x\approx 5$ MeV.} 
  \label{fig-density}
\end{figure}

\section{Similarities and differences between the two model spaces}
\label{sec-simila}

As was already noticed in the calculations carried out by Harder {\it et
al.}~\cite{harder97} and in the calculations presented in section
\ref{sec-fit-procedure}, 
there are a number of properties in the
Pt nuclei, such as excitation energies and $B(E2)$ values 
for the set of levels below $\approx$
1.5 MeV, that do not seem to be very sensitive to the use of a smaller model space as compared
to the use of a much larger model space incorporating both regular and particle-hole excited intruder
configurations, explicitly.
This can partly be expected because
precisely those observables - excitation energies, 
$B(E2;2^+_1 \rightarrow 0^+_1)$, $B(E2;2^+_2 \rightarrow 0^+_1)$, 
and $B(E2;2^+_2 \rightarrow 2^+_1)$ - 
when experimentally known, were
used in the fitting procedure in both the IBM-CM and the ECQF calculations.

\subsection{The energy spectra at high excitation energy: comparing the IBM-CM and ECQF}
\label{sec-high}

When comparing the IBM-CM calculations (which is using a model space
that contains both, $N$ boson and $N+2$ boson configurations) with the
ECQF calculations (which is using a model space with N bosons only),
it is unavoidable that moving up in excitation energy, at some point,
clear-cut differences in the density of states with given $J^{\pi}$
value should show up (the full dimension in the IBM-CM more than
doubles the one of the ECQF).  It is interesting to make the
comparison between both. In Fig.~\ref{fig-density}, we
illustrate this in the specific case of $^{184}$Pt (which
is typical for all nearby nuclei), which is situated in a region where
a number of unperturbed configurations containing $N$ and $N+2$
bosons, respectively, appear at about the same unperturbed energy and
thus start a complex crossing pattern in the IBM-CM approach. 
In Fig.~\ref{fig-density}, one observes that up to an excitation energy
of about $2$ MeV, there are no obvious differences between the IBM-CM
and the ECQF theoretical results. Above this energy, different patterns are
showing up. In particular in the IBM-CM one observes a rather smooth
distribution of levels, while in the ECQF the energy levels appear more
separated in blocks spaced by about $0.5$ MeV. This separation in
blocks accentuates when increasing the value of the angular momentum.
This turns out to be an indication that one is running out of model space if
one restricts to $[N]$ configurations only.

\subsection{Eliminating the N+2 configurations: a mapping procedure}
\label{sec-elimina}

We point out that the very close resemblance of a part of the energy spectrum
(restricting to excitation energies below $\approx$ 1.5 MeV) and
corresponding $B(E2)$ values,  
can most probably be understood from a mapping of the larger model space, used in the IBM-CM in which
configurations with both $N$ and $N+2$ components are considered, onto the smaller space, used in the ECQF 
formulation, in which only the
$N$ boson configurations are kept. Thereby, an effective Hamiltonian acting in the smaller space is defined through
the mapping of the lowest energy eigenvalues and similarly for the corresponding model transition operators.
We can describe the wave functions in the IBM-CM as follows
\begin{eqnarray}
\Psi(k,JM) &=& \sum_{i} a^{k}_i(J;N) \psi((sd)^{N}_{i};JM) 
\nonumber\\
&+& \
\sum_{j} b^{k}_j(J;N+2)\psi((sd)^{N+2}_{j};JM)~.
\end{eqnarray}
We can also consider a limited number of basis states, {\it i.e.,}~only the components with $N$ bosons. This basis
spans the smaller model space and the corresponding  model wave function can be expressed as
\begin{equation}
\Psi'(k,JM) = \sum_{l} a'^{k}_{l}(J;N) \psi((sd)^{N}_{l};JM \rangle ~,
\end{equation}
{\it i.e.}, that part of the ``true'' wave function that lies within
the small model space. 
If we then require that the
lowest energy eigenvalues $E_k(JM)$, for the large space, 
be reproduced exactly within the much smaller model space, 
one can determine an effective Hamiltonian acting in the
reduced model space containing only configurations with N bosons by a mapping procedure
\begin{equation}
\langle \Psi'(k,JM)\mid \hat {H}^{eff.} \mid \Psi'(k,JM) \rangle = E(k,JM) ~.
\end{equation}
Using the projection operator $\hat{\bf P} = \sum_{i\subset N} \mid (sd)^{N}_{i};JM \rangle \langle (sd)^{N}_{i};JM \mid$,
which projects onto the model space containing only N-boson
configurations and the operator $\hat{\bf Q} =
\sum_{j\subset {N+2}} \mid (sd)^{N+2}_{j};JM \rangle \langle (sd)^{N+2}_{j};JM \mid$, projecting onto the N+2-boson 
model configurations, an effective Hamiltonian acting in the reduced model
space can be constructed \cite{brussaard77} as
\begin{equation}
\hat{H}^{eff.} = \hat{H} + \hat{H} \frac{\hat{\bf Q}}{E(k,JM) - \hat{H}^0} \hat{H} + ....
\label{v_eff}
\end{equation}

\begin{figure}[hbt]
  \centering
  \epsfig{file=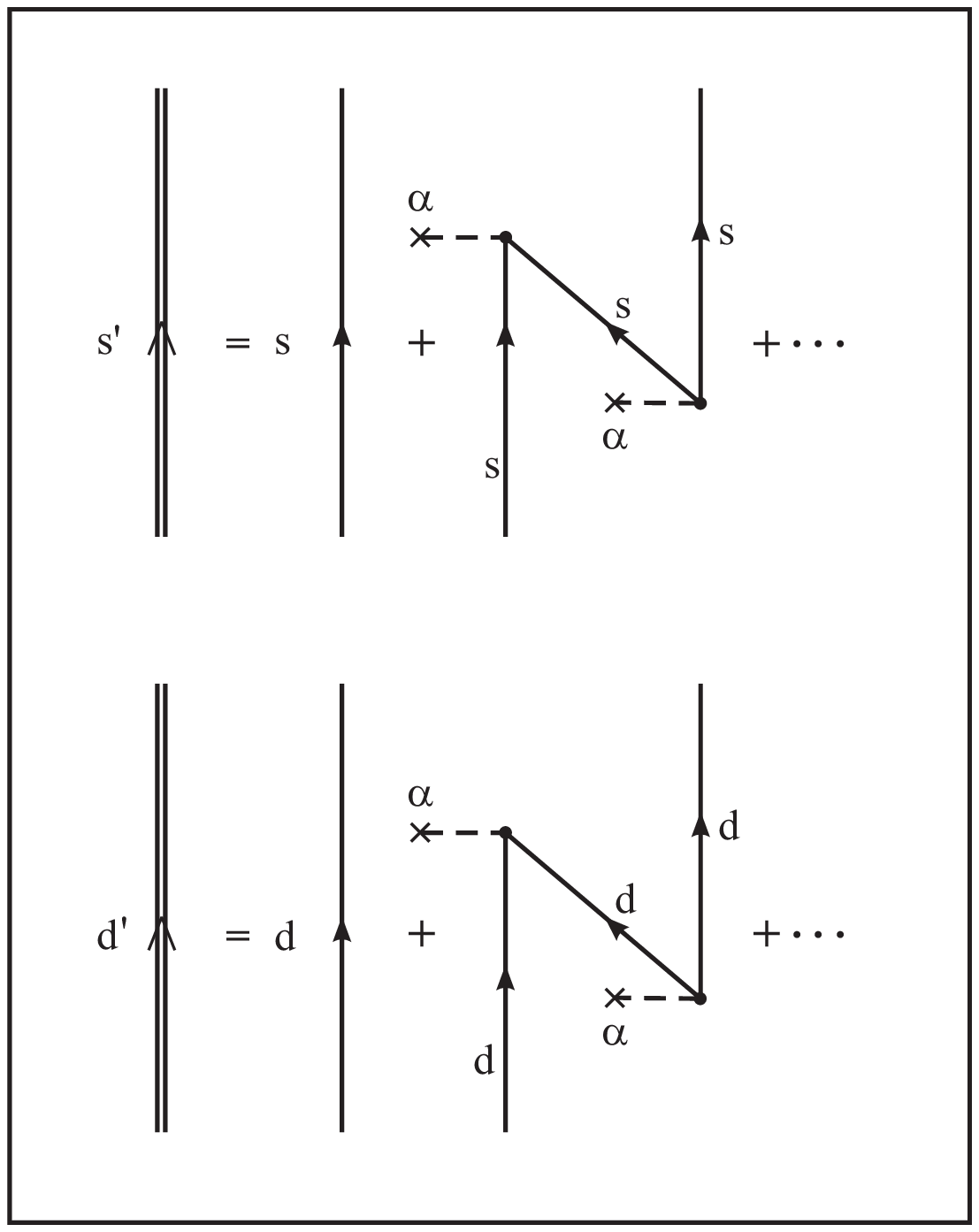,width=6cm} 
  \caption{Schematic diagram representing the effect of eliminating
    the N+2 space on the single particle boson energies.}
  \label{fig-renor-single}
\end{figure}

Although this formal procedure cannot easily be carried out starting from the Hamiltonian used in the IBM-CM
model space to construct the effective Hamiltonian acting in the reduced model space with N bosons only, it seems 
possible to show numerically that the procedure will work. Indeed, the low-lying levels calculated in
the IBM-CM model space up to $E_x \approx 1.5$ MeV match up very well with the corresponding levels using
the ECQF calculation within the N-boson model space, only. Considering the lowest-order diagrams that
follow from the expansion in Eq.~(\ref{v_eff}), 
it can easily be seen that eliminating the N+2 components implies
changes in both the boson single-particle energies $\varepsilon_s$ and
$\varepsilon_d$ (see Fig.~\ref{fig-renor-single}), as well as
in the parameters $\kappa$, $\kappa'$, and $\chi$ appearing in the interaction part
of the Hamiltonian. 
Work is
in progress to try to carry out the mapping for the most simple terms
appearing in the IBM-CM Hamiltonian.

\subsection{Observables and sensitivity to the enlarged model space}
\label{sec-sensitiv}

The idea of comparing truncated model spaces with much larger model spaces and
the fact that a number of observables obtained in these model spaces might still 
turn out to be very similar when comparing to experimental data was illustrated  
a long time ago by Cohen, Lawson and Soper \cite{cohen66,law80}. 
By theoretically constructing a set of nuclei, called the Pseudonium nuclei $^{40-48}$Ps, starting from a
model space consisting of two degenerate $1d_{3/2}$ and $1f_{7/2}$ single-particle orbitals,
containing between 4 and 12 neutrons and a given two-body interaction (a Yukawa force was used),
they showed that these energy spectra interpreted as pseudo-data, could be fitted very well
using a much restricted model space consisting of the $1f_{7/2}$ orbital only, now containing
between 0 and 8 neutrons. It turned out that the effective interaction matrix elements, fitted
to the Pseudonium nuclei, was corresponding to quite a different interaction than
the force used at first in the larger model space. They moreover showed that other observables,
such as $B(E2)$ values for the strongest transitions, were very similar even though the wave functions
were very different. A different set of Pseudonium nuclei were constructed making
use of a model space consisting of two degenerate $1p_{1/2}$ and $1d_{3/2}$  single-particle states
that could contain both protons and neutrons, up to maximally 12 nucleons. Very much the same
conclusion resulted when analyzing corresponding states using a model space consisting of the $1d_{3/2}$ 
orbital only \cite{law67}. In the latter study, it was pointed out that quadrupole moments seemed 
to be a better observable to probe differences, with particularly chosen transfer reactions, becoming
highly sensitive to the use of different model spaces. It shows that a number of observables
(excitation energies, $B(E2)$,...) are very insensitive to
configuration mixing arising from 
the excitation of zero-coupled
pairs out of the closed shell. This may turn out to be the same underlying mechanism in 
our present IBM-CM study.

This study of the actual wave function content and the way to test it
has, more recently, been explored, 
{\it e.g.}, in the study of the nucleus $^{40}$Ca \cite{caurier07}. It
turns out that the $0^+$ ground state 
only consists of 65\% 
closed sd shell (or 0p-0h) and exhibits 29\% of 2p-2h excitations
out of the $2s_{1/2},1d_{3/2}$ normally filled orbitals into the $1f_{7/2},2p_{3/2},1f_{5/2},2p_{1/2}$
higher-lying orbitals, even containing up to 5\% of 4p-4h excitations. This large model space is effectively needed
in order to describe the higher-lying strongly deformed bands and
superdeformation as experimentally observed in $^{40}$Ca. 
A study of isotopic shifts in the even-even Ca nuclei, moving from
$A=40$ to $A=48$ could be reproduced well by including  
explicitly, mp-nh excitations across the $Z=20$, $N=20$ "closed"
shell, using a slightly smaller model 
space than the one used before \cite{caurier01}. This points out that
one can indeed find observables 
that are sensitive to the important components of the wave function
and thus be able to discriminate between 
various approaches that give quite similar results when restricting to
a subset of data, only. 

Therefore, in a forthcoming paper, we shall compare the results of
both, the IBM-CM and ECQF calculations with a data set that encompasses
besides excitation energies and $B(E2)$ values also 
quadrupole moments, $g(2_1^+)$
values, $\alpha$-decay hindrance factors, isotopic shifts, $\rho^2$
values extracted from E0 transitions, as well as results derived from
properties characterizing the energy spectra and E0 transitions in the
adjacent odd-mass nuclei. Transfer reactions are most probably one of
the best tests to explore the detailed composition of the nuclear wave
function. They have been shown to be able to detect the presence of
core-excited configurations in different mass regions. Such a
comparison should allow to compare
the nuclear wavefunctions derived in the smaller N space with the
wavefunctions containing both N and N+2 configurations (inclusion of
particle-hole excitations, explicitly).

\section{Conclusions}
\label{sec-conclu}
In the present study, we have analyzed certain aspects related to the role of intruder excitations in the
Pt isotopes. To start with, we have presented the 
experimental knowledge in this mass region on energy spectra and absolute $B(E2)$
values, however restricting to an excitation energy up to $1.5$ MeV. 
Next, we have performed a set of IBM-CM calculations for the isotopes $^{172-194}$Pt
and compared them with previous ECQF calculations, noticing 
that for the subset of states below $\approx 1.5$ MeV, both descriptions are roughly comparable.
However, if one extends both calculations up to 5 MeV, differences in the distribution
of the theoretical levels start to show up. 
We have suggested that the similarities between both model spaces 
can be explained through an appropriate mapping with a low-energy cut-off in the excitation
energy of $\approx 1.5$ MeV. This mapping may shed light on the way to extract the parameters 
that appear in the smaller model space (which amounts to four in the ECQF study) 
starting from the larger model space, containing in the present IBM-CM approach eight free parameters.
We have also made reference to earlier studies by Cohen, Lawson and Soper, 
carried out within the context of the nuclear shell-model,
where it was shown that calculated energy spectra and other
observables, using a given model space 
and two-body interaction in the case of strong mixing, can be fitted very well using a restricted
model space and an effective or renormalized interaction. 
They showed the insensitivity of a number of observables (such as
excitation energies and $B(E2)$ values)
to strong configuration mixing arising from the excitation of $0^+$ coupled
pairs out of a closed shell. 
These studies stressed the need to detect (i) all states (complete spectroscopy) as well as, 
(ii) the importance to study  
specific observables in order to probe the nuclear wavefunction in greater detail.

The fact that for the neutron-deficient Pb and Hg nuclei, a fit using the smaller model space
cannot be obtained (in particular the low-lying $0^+_2$ energies)~\cite{cutcham05} is particularly
relevant in this respect. It shows that in the case of weak mixing between the regular low-lying
configurations (N-space) and particle-hole (p-h) pair excitations
across $Z=82$ closed shell (N+2,...~space), the
intruder states are clearly visible and exhibit a distinct energy systematics. This points out the
need to consider a model space containing both regular and p-h pair configurations. However, in 
situations where the p-h pair excitations appear at nearly the same unperturbed energy as the regular
excitations and strong mixing follows, the observed excitation energies do not show any more such a
distinct separation. Still, the excitation energies and electric quadrupole properties for states
below $E_x \approx$ 1.5 MeV are very similar using the larger 
and the smaller model space,
excluding p-h degrees of freedom, explicitely (section~\ref{sec-fit-compare} and 
section~\ref{sec-fit-intr}).
This is a most interesting result, in view of the highly-different wave functions that enter the
calculations and may be understood constructing the mapping referred to before. 
Moreover, the arguments put forward in the study of the Pseudonium nuclei by Cohen, Lawson and Soper
of concealed configuration mixing of $0^+$ coupled pairs, are interesting in the prospect of
accomplishing a mapping. Work is in progress on this important issue.   
Moreover, in a forthcoming paper, we shall present  
the results of a comparison covering an as complete as possible data set (encompassing also,  
\textit {e.g.}, $\alpha$-decay hindrance
factors, $g(2^+_1)$ factors, isotopic shifts, E0 decay properties, as well as   
properties in the spectra of neighboring odd-mass Pt and Au nuclei) with the two theoretical
approaches, {\it i.e.,}~considering a reduced model space versus a
larger model space including particle-hole excitations explicitly,
within the Interacting Boson Model context. Transfer reactions may prove to be very effective in
showing the detailed structure of the nuclear wave functions.

\section{Acknowledgment}
We thank M. Huyse, P.Van Duppen, and P.Van Isacker 
for continuous interest in this research topic and 
J.L.~Wood for stimulating discussions in various stages of this work.
Financial support from the ``FWO-Vlaanderen'' (KH and JEGR) is acknowledged. This research was
also performed in the framework of the BriX network (IAP P6/23) funded by the 'InterUniversity
Attraction Poles Programme - Belgian State-Belgian Science Policy'.
This work has also been partially supported
by the Spanish Ministerio de Ciencia e Innovaci\'on and by the European
regional development fund (FEDER) under projects number 
FPA2006-13807-C02-02 and FPA2007-63074, by Junta de
Andaluc\'{\i}a under projects FQM318, and P07-FQM-02962 and by the
Spanish Consolider-Ingenio 2010 Programme CPAN (CSD2007-00042).

\appendix
\section{Appendix: Relative $B(E2)$ values in $^{180-194}$Pt}
\label{ap-A}

Since the number of absolute $B(E2)$ values known in the $^{180-194}$Pt nuclei is quite restricted, we also
cover relative $B(E2)$ values. Extracting those relative values, 
one needs 
detailed information on the $\gamma$-ray
intensities for the transitions originating from a given initial state. In order to do so, we have made use
of the adopted values given in the most recent volume of the
Nuclear Data Sheets (NDS), covering the mass range
from $A=180$ up to $A=194$ and are taken from \cite{Wu03} ($A=180$),
\cite{Bagl03} ($A=186$), \cite{Sing02} ($A=188$), 
 \cite{Sing03} ($A=190$), \cite{Bagl98} ($A=192$) and \cite{Sing06} ($A=194$). In a number of cases, {\it i.e.,} when 
the latest issue of NDS was published more than 10 years ago, we have taken data from recent literature. This was 
the case for $A=182$ (\cite{Sing95b} dates from 1995), in which case the necessary data were taken from a study of
Davidson {\it et al.}~\cite{davidson99} and also for $A=184$ with the NDS dating to 1989 \cite{Fire89}. In the
latter case, we have used the update on $^{184}$Pt, kindly supplied by Baglin \cite{baglin09}, information which 
will soon be published in a forthcoming issue of NDS. For $A=192$, we have, for the transitions decaying from the
$2^+_3$ level, taken the data from McCutchan {\it et.al.} \cite{cutcham08}.   

We have taken from these data the transition energies, intensities and corresponding errors. In the table that follows,
we also give the evaluated information on multipolarities in
situations where one can expect that besides $E2$ 
transitions, $M1$ transitions and, in situations where $J_i=J_f$, also $E0$ transitions can contribute. Thereby we
follow the conventions used by the evaluators of NDS: (...) stands for very strong indications for the multipolarities
given between parenthesis, [   ] stands for multipolarities deduced from information on the decay scheme, even
compelling, but not measured, (Q) or Q stands for a transitions with quadrupole multipolarity, but no unambigous
parity information known. In a number of cases, $\delta(E2/M1)$ mixing ratios have been measured and are given.
In a number of transitions that have been taken as a reference transition to derive the relative $B(E2)$ values,
but might contain $M1$ admixtures, we explicitely mention if a mixing
ratio is know. 
In a number of situations such as the $2^+_2 \rightarrow 2^+_1$ transition, the $E2$ assignment in NDS,
even with the mixing ratio unknown, is made on the basis of measured conversion coefficients, excluding important
$M1$ admixtures. For all nuclei considered here, we have assumed this particular transition to be $E2$, so the 
numbers deduced from that may carry a given ``uncertainty'' with them. Something quite similar occurs for the
 $3^+_1 \rightarrow 2_2$ transition, which has (if present) been taken as the reference transition when deriving
relative $B(E2)$ values. In a number of cases, the mixing ratio helps substantially in deciding to fix a reference
transition in order to derive relative $B(E2)$ values.

Using the criteria discussed before, we have extracted the experimental relative $B(E2)$ values (with error).
The expression used is given by
\begin{equation}
\mbox{B(E2)}= 100 \times \left (\frac{\mbox{I}_\gamma}{\mbox{I}_\gamma^{ref}}\right )\times
\left(\frac{\mbox{E}_\gamma^{ref}}{\mbox{E}_\gamma }\right)^5 \times \left (
\frac{1+\frac{1}{\delta^2_{ref}}}{1+\frac{1}{\delta^2}}\right),  
\end{equation} 
where {\it ref} stands for the reference transition.

The related errors are given by the expression
\begin{eqnarray}
\nonumber&&\Delta(\mbox{B(E2)})= \mbox{B(E2)}\\
&&\times\sqrt{\
\left (\frac{\Delta(\mbox{I}_\gamma)}{\mbox{I}_\gamma}\right)^2+
\left(\frac{\Delta(\mbox{I}_\gamma^{ref})}{\mbox{I}_\gamma^{ref}}\right)^2 
+4 \left(\frac{\Delta(\delta)}{\delta}\right)^2
+4 \left(\frac{\Delta(\delta_{ref})}{\delta_{ref}}\right)^2
      },
\end{eqnarray} 
where $\Delta()$ stands for the absolute error.

Here, the relative errors for both the reference transition, the one
that is taken as the relative transition, as well as
the relative errors on mixing ratios (again for the reference
transition and the one that is taken as the relative transition), 
if the latter are available, have to be taken into account in a
quadratic way. Note that we did not include the contribution from the
error of the gamma energies because they are negligible.
\begin{table}
\begin{center}
\begin{tabular}{|c|c|c|c|c|c||c|c|c|}
\hline
   &Transition
   &E$_\gamma$&I$_\gamma$&Mult.&$\delta$&Exp.& IBM-CM & ECQF\\
 & &(keV) &&&&&&\\
\hline 
$^{180}$Pt&$2_2^+\rightarrow 0_2^+$&199 &5(2)&&&{ 630(250)}& { 84}   &{ 120}\\	
         &$2_2^+\rightarrow 4_1^+$&207 &0.9(5)& &&{ 26(15)} & { 13}   &{ 20} \\	
         &$2_2^+\rightarrow 2_1^+$&524 &100(2)&E0+E2&$<$-11&{ 100 }   & { 100}  &{ 100}   \\	
         &$2_2^+\rightarrow 0_1^+$&678 &35(3)&&&{ 9.7(9)}   & { 4.5}  &{ 6}  \\	 
\cline{2-9}
         &$2_3^+\rightarrow 2_2^+$&184 &1.1(6) & & &  { $<$304}   & { 158}   &{ 158}   \\ 
         &$2_3^+\rightarrow 0_2^+$&383 &14(2) & & &  { 100}    & { 100}  &{ 100 }  \\ 
         &$2_3^+\rightarrow 4_1^+$&451 &18(2) & & &  { 57(10)} & { 44}   &{ 50} \\	
         &$2_3^+\rightarrow 2_1^+$&708 &15(2) &E0+M1+E2 &2.0$^{+3.6}_{-1.1} $ & 4.0$^{+1.7}_{-2.2}$  & { 3.2}&{ 0.5}  \\	
         &$2_3^+\rightarrow 0_1^+$&861 &100(2) & & &  { 12.4(1.8)}  & { 13}   &{ 10} \\	
\cline{2-9}
         &$3_1^+\rightarrow 2_2^+$&286 &3.5(7) & $^*$&Unknown &  { 100}    & { 100}  &{ 100}   \\	
         &$3_1^+\rightarrow 4_1^+$&552 &13(2) & & &  { $<$14}    & { 17} &{ 17} \\	
         &$3_1^+\rightarrow 2_1^+$&889 &100(2) &E2 &$<$-13 &  { 16(3)} & { 11}   &{ 9}   \\	
\cline{2-9}
         &$4_2^+\rightarrow 2_2^+$&372 &15(4) & & &  { 100}    & { 100}  &{ 100}   \\	
         &$4_2^+\rightarrow 4_1^+$&639 &100(13)& & &  { 45(13)} & { 35}   &{ 25} \\	
         &$4_2^+\rightarrow 2_1^+$&896 &17(8) & & &  { 1.4(8)} & { 0.1}  &{ 0.2}  \\	
\cline{2-9}
         &$0_3^+\rightarrow 2_2^+$&500 &13(8) & & &  { 100}    & { 100}  &{ 100}   \\	
         &$0_3^+\rightarrow 2_1^+$&1024 &100(10) & & &  { 21(13)} & { 1.4}  &{ 0.33}  \\	
\cline{2-9}
         &$4_3^+\rightarrow 2_3^+$&387 &70(40) & & &  { 100}    & { 100}  &{ 100}   \\	
         &$4_3^+\rightarrow 2_2^+$&572 &$\approx$ 21 & & &  { $\approx$4}   & { 5.1}  &{ 0.5}  \\	
         &$4_3^+\rightarrow 4_1^+$&837 &27(7) & & &  { $<$0.8 }    & { 0.8}&{ 0.7}  \\	
         &$4_3^+\rightarrow 2_1^+$&1095 &100(20) & & &  { 0.8(5)} & { 4.1}  &{ 0.9}  \\	
\cline{2-9}
         &$5_1^+\rightarrow 3_1^+$&353 &19(5) & & &  { 100 }   & { 100}  &{ 100}   \\	
         &$5_1^+\rightarrow 4_1^+$&905 &100(9) &(M1+E2) & &  { $<$5  }   & { 4.6}&{ 3.3}  \\	  
\hline
$^{182}$Pt
          &$ 2_2^+\rightarrow 2_1^+$&512 &23.3(26) &E0+(M1)+E2$^*$& Unknown &  { 100} & { 100 } &{ 100}\\     
          &$ 2_2^+\rightarrow 0_1^+$&667 &8.0(10) & & & { 9.2(15)}    & { 12   }&{ 6.4}    \\           
\cline{2-9}
          &$ 2_3^+\rightarrow 0_2^+$&356 &2.6(7) & & & { 100}    & { 100   }&{ 100}    \\           
          &$ 2_3^+\rightarrow 4_1^+$&436 &3.4(3) & & & { 47(13)}  & { 25    }&{ 49}  \\   
          &$ 2_3^+\rightarrow 2_1^+$&701 &1.3(3) &E0+M1+E2 &0.7$^{+1.0}_{-0.3}$ & {0.6$^{+1.2}_{-0.4}$} & { 2.8   }&{ 0.7}   \\  	
          &$ 2_3^+\rightarrow 0_1^+$&855 &15.0(5) & & & { 7.2(19)}   & { 3.2   }&{ 5.3}   \\          
\cline{2-9}
          &$ 3_1^+\rightarrow 4_1^+$&523 &2.5(3) &[M1,E2] & & { $<$146} & { 156   }&{ 189}   \\  	
          &$ 3_1^+\rightarrow 2_1^+$&787 &13.7(19) &(M1+E2) &$>$5 & { 100} & { 100  }&{ 100}   \\  	

\hline 
\end{tabular}
\end{center}
  \caption{Comparisons of the experimental relative $B(E2)$ values  with
    the IBM-CM Hamiltonian and the ECQF results \cite{cutcham05}.
>From left to right we give: isotope, transition, $\gamma$-ray energy,
intensity of the transition, multipolarity, $\delta$ value,
experimental relative $B(E2)$ value, IBM-CM calculation and ECQF calculation. 
Data are taken from 
\cite{Wu03,
Bagl03,Sing02,Sing03,Bagl98,Sing06}, C.~Baglin \cite{baglin09}, complemented with
references \cite{davidson99,cutcham08}.
\newline
$^*$ Pure E2 transition assumed.} 
  \label{tab-be2-1}
\end{table}

\begin{table}
\begin{center}
\begin{tabular}{|c|c|c|c|c|c||c|c|c|}
\hline
   &Transition
   &E$_\gamma$&I$_\gamma$&Mult.&$\delta$&Exp.& IBM-CM & ECQF\\
 & &(keV) &&&&&&\\
\hline 
$^{184}$Pt&$ 2_2^+\rightarrow 2_1^+$&486 &100(10) &(E0)+E2+M1$^*$ &Unknown & { 100}    & { 100}   &{ 100} \\          
         &$ 2_2^+\rightarrow 0_1^+$&649 &50(8) & & & {  61(21)}    & { 83}   &{ 87} \\
\cline{2-9}
         &$ 2_3^+\rightarrow 2_2^+$&195 &2.6(8) & (E0)+M1+E2& & { $<$415 }    & { 122}   &{ 110}   \\          
         &$ 2_3^+\rightarrow 0_2^+$&352 &12.0(18) &[E2] & & { 100   }    & { 100}   &{ 100}   \\          
         &$ 2_3^+\rightarrow 4_1^+$&408 &12.7(19) & & & { 51 (11)}   & { 55 }   &{ 48} \\         
         &$ 2_3^+\rightarrow 2_1^+$&681 &13.1(20) &E0+M1+E2 &-1.2$^{+0.5}_{-3.5}$ & [1.0,4.9]& 0.8 &{ 0.3}  \\         
         &$ 2_3^+\rightarrow 0_1^+$&844 &100(15) & & & { 10.5 (22)}& { 10.3}  &{ 3.6}  \\  
\cline{2-9}
         &$ 3_1^+\rightarrow 2_2^+$&291 &3.6(11) &$^*$ & Unknown & { 100 }& { 100}   &{ 100}  \\    	   
         &$ 3_1^+\rightarrow 4_1^+$&504 &15.7(25) & & & { $<$28 }& { 16}   &{ 18}  \\    	   
         &$ 3_1^+\rightarrow 2_1^+$&777 &100(10) &(M1)+E2 &$>$4 &  { 19(1) }& { 10}   &{ 7.9}  \\    	   
\cline{2-9}
         &$ 2_4^+\rightarrow 2_2^+$&524 &8.5(26) &E0+M1+E2$^*$ & Unknown & { 100    }   & { 100}   &{ 100}   \\         
         &$ 2_4^+\rightarrow 0_2^+$&681 &12(3) & & & { 38(15)  }   & { 0.8}   &{ 59} \\         
         &$ 2_4^+\rightarrow 0_1^+$&1172 &22(7) & & & { 4.6 (20)} & { 0.1}   &{ 2.1}  \\       
\cline{2-9}
         &$ 4_2^+\rightarrow 2_2^+$&379 &39(6) & & & { 100 }& { 100}   &{ 100}  \\    	   
         &$ 4_2^+\rightarrow 4_1^+$&592 &100(14) &(E0)+M1+E2 &$\leq$-1.2 & { 21(5) }& { 35}   &{ 30}  \\    	   
\cline{2-9}
         &$ 4_3^+\rightarrow 3_1^+$&294 &7.7(23) & & & { $<$62    }   & { 104}   &{ 30}   \\         
         &$ 4_3^+\rightarrow 2_3^+$&390 &51(7) & & & { 100    }   & { 100}   &{ 100}   \\         
         &$ 4_3^+\rightarrow 2_2^+$&586 &18(6) & & & { 4.6 (17)   } & { 8.7}   &{ 0.3}  \\       
         &$ 4_3^+\rightarrow 4_1^+$&798 &74(12) &E0+M1+E2 &1.1(3) & { 2.2 (1.3) }& { 0.4}   &{ 0.5}  \\      
         &$ 4_3^+\rightarrow 2_1^+$&1071 &100(14) & & & { 1.26 (25) }& { 2.7}   &{ 0.7}  \\    	   
\cline{2-9}
         &$ 3_2^+\rightarrow 2_4^+$&297 & 29(8)&$^*$ & Unknown& { 100 }& { 100}   &{ 100}  \\    	   
         &$ 3_2^+\rightarrow 3_1^+$&530 &$<$1.4 &E0+M1+E2 & & { $<$0.27 }& { 0}   &{ 0}  \\    	   
         &$ 3_2^+\rightarrow 2_3^+$&626 &24(7) & & & { $<$2.0 }& { 0.03}   &{ 1.7}  \\    	   
         &$ 3_2^+\rightarrow 2_2^+$&821 & 44(13)& & & { $<$0.9 }& { 0.4}   &{ 0.6}  \\    	   
         &$ 3_2^+\rightarrow 4_1^+$&1034 &60(18) & & & {$<$ 0.40 }& { 0.02}   &{ 0.08}  \\    	   
         &$ 3_2^+\rightarrow 2_1^+$&1307 &100(15) & & & {$<$ 0.21 }& { 0.03}   &{ 0.04}  \\    	   
\hline 
\end{tabular}
\end{center}
  \caption{See caption of Table \ref{tab-be2-1}.}
  \label{tab-be2-2}
\end{table}

\begin{table}
\begin{center}
\begin{tabular}{|c|c|c|c|c|c||c|c|c|}
\hline
   &Transition
   &E$_\gamma$&I$_\gamma$&Mult.&$\delta$&Exp.& IBM-CM & ECQF\\
 & &(keV) &&&&&&\\
\hline 
$^{186}$Pt&$ 2_2^+\rightarrow 2_1^+$&416 &100(4) &M1+E2 &19$^{+20}_{-8}$ & { 100}    & { 100} &{ 100}   \\ 	
         &$ 2_2^+\rightarrow 0_1^+$&607 &62(15) & & & { 9.4(23)}   & {   5} &{ 6.3}  \\	
\cline{2-9}
         &$ 2_3^+\rightarrow 4_1^+$&308 &7.7(6) & & & { 69(8)}  & {  38} &{ 39} \\	
         &$ 2_3^+\rightarrow 0_2^+$&327 &15.1(12) & & & { 100}    & { 100} &{ 100}   \\	
         &$ 2_3^+\rightarrow 0_1^+$&799 &100(12) & & & { 7.6(11)}   & { 3.8} &{ 2.2}  \\	
\cline{2-9}
         &$ 3_1^+\rightarrow 2_2^+$&349 &12.4(8) &M1+E2 &2.8(3) & { 100}    & { 100} &{ 100}   \\	
         &$ 3_1^+\rightarrow 4_1^+$&466 &12.4(6) &(M1+E2) & 0.42(7)& { 4.0(16)}  & {  24} &{ 19} \\	
         &                         & & & & 3.8(9)& { 25(13)}  & {  24} &{ 19} \\	
         &$ 3_1^+\rightarrow 2_1^+$&765 &100(5) &M1+E2 &16$^{+4}_{-3}$ & { 18(4)}  & {   9.0} &{ 6.7}  \\	
\cline{2-9}
         &$ 4_2^+\rightarrow 2_2^+$&384 &63(5) & & & { 100}    & { 100} &{ 100}   \\	
         &$ 4_2^+\rightarrow 4_1^+$&501 &100(16) &M1+E2 &-0.85(9) & { 18(5)}  & {  37} &{ 39}  \\	
         &$ 4_2^+\rightarrow 2_1^+$&800 &79(16) & & & { 3.2(7)}   & { 0.4} &{ 0.3}  \\	
\cline{2-9}
         &$ 2_4^+\rightarrow 0_2^+$&704 &43(5) & & & { 100}    & { 100} &{ 100}   \\	
         &$ 2_4^+\rightarrow 2_1^+$&985 &100(10) &M1+E2 & -0.12(6)& { 0.6(6)}  & {  41} &{ 49} \\	
         &&& &M1+E2 & 3.2(8)& { 40(21)}  & {  41} &{ 49} \\	
         &$ 2_4^+\rightarrow 0_1^+$&1176 &48(5) & & & { 8.6(13)}   & { 3.5} &{ 2.4}  \\	   
\cline{2-9}
         &$ 4_3^+\rightarrow 3_1^+$&266 &19.1(18) & & & { $<$440}    & { 115} &{ 25}   \\	
         &$ 4_3^+\rightarrow 2_3^+$&424 &45(5) & & & { 100}    & { 100} &{ 100}   \\	
         &$ 4_3^+\rightarrow 4_1^+$&732 &82(5) &E0+M1+E2 & & { $<$12}  & { 2.3} &{ 0.3}  \\	
         &$ 4_3^+\rightarrow 2_1^+$&1031 &100(5) & & & { 2.6(3)} & { 1.4} &{ 0.6}  \\	
\cline{2-9}
         &$ 5_1^+\rightarrow 3_1^+$&406 &22(7) & & & { 100}    & { 100} &{ 100}   \\	
         &$ 5_1^+\rightarrow 4_1^+$&872 &100(7) & & & { $<$10}  & { 4.1} &{ 3.3}  \\	
\hline 
\end{tabular}
\end{center}
  \caption{See caption of Table \ref{tab-be2-1}.}
  \label{tab-be2-3}
\end{table}

\begin{table}
\begin{center}
\begin{tabular}{|c|c|c|c|c|c||c|c|c|}
\hline
   &Transition
   &E$_\gamma$&I$_\gamma$&Mult.&$\delta$&Exp.& IBM-CM & ECQF\\
 & &(keV) &&&&&&\\
\hline 
$^{188}$Pt&$ 2_2^+\rightarrow 2_1^+$&340 &100(4) &E2$^*$ &Unknown & { 100   }& {  100} &{ 100}   \\	
         &$ 2_2^+\rightarrow 0_1^+$&605 &66(3) & & & { 3.7(2)}& {  0.5} &{ 2.2}  \\	
\cline{2-9}
         &$ 0_2^+\rightarrow 2_2^+$&193 &2.9(7) & & & { 100   }& {  100} &{ 100}   \\	
         &$ 0_2^+\rightarrow 2_1^+$&533 &100(4) & & & { 21(5) }& {  17}  &{ 53} \\	
\cline{2-9}
         &$ 3_1^+\rightarrow 2_2^+$&331 &63(3) &E2$^*$ &Unknown & { 100}& {  100} &{ 100}  \\	  
         &$ 3_1^+\rightarrow 2_1^+$&671 &100(4) & & & { $<$5}& {  0.84} &{ 3.3}  \\	  
\cline{2-9}
         &$ 4_2^+\rightarrow 4_1^+$&415 &45(4) &M1+(E2) & & { $<$92 }& {  83}  &{ 69} \\	
         &$ 4_2^+\rightarrow 2_2^+$&479 &100(7) & & & { 100   }& {  100} &{ 100}   \\	
         &$ 4_2^+\rightarrow 2_1^+$&820 &21(2) &(Q) & & { 1.43(17) }& {0.4 } &{0.14 }   \\	
\cline{2-9}
         &$ 2_3^+\rightarrow 0_2^+$&316 &19(2) & & & { 100   }& {  100} &{ 100}   \\	
         &$ 2_3^+\rightarrow 4_1^+$&444 &22(2) & & & { 21(3) }& {  17}  &{ 24} \\	
         &$ 2_3^+\rightarrow 2_1^+$&849 &14(5) &E0+M1+E2 & & { $<$0.5}& {  0.3} &{ 0.07}  \\	
         &$ 2_3^+\rightarrow 0_1^+$&1115 &100(4) & & & { 1.0(1)}& {0.1  } &{ 0.8}  \\	  
\cline{2-9}
         &$ 2_4^+\rightarrow 2_3^+$&198 &7.1(24) &$^*$ & Unknown& { 100}& {  100} &{ 100}  \\	  
         &$ 2_4^+\rightarrow 3_1^+$&376 &17(3) &M1+E2 &1.3(3) & { 6(4)}& {  39} &{ 1240}  \\	  
         &$ 2_4^+\rightarrow 4_1^+$&642 &22(4) & & & { 0.9(3)}& {  0.01} &{ 1.3}  \\	  
         &$ 2_4^+\rightarrow 2_2^+$&707 &31(4) &E0+M1+E2 & & { $<$0.8}& {  0.3} &{ 16}  \\	  
         &$ 2_4^+\rightarrow 2_1^+$&1047 &65(7) &E0+M1+E2 & & { $<$0.2}& {  0.05} &{ 0.7}  \\	  
         &$ 2_4^+\rightarrow 0_1^+$&1313 &100(7) & & & { 0.11(4)}& {  0.0} &{ 0.0}  \\	  
\hline 
$^{190}$Pt&$ 2_2^+\rightarrow 2_1^+$&302 &100(3) &E2$^*$ &Unknown & { 100      }& { 100} &{ 100}   \\	  
         &$ 2_2^+\rightarrow 0_1^+$&598 &40(3) & & & { 1.3(1)   }& { 0}   &{ 1.0}  \\	  
\cline{2-9}
         &$ 3_1^+\rightarrow 4_1^+$&180 &3.1(3) &M1+E2 &3$^{+2}_{-1}$ & { 49$^{+9}_{-10}$   }& { 43}  &{ 36} \\	  
         &$ 3_1^+\rightarrow 2_2^+$&319 &100(5) &E2$^*$ &Unknown & { 100      }& { 100} &{ 100}   \\	  
         &$ 3_1^+\rightarrow 2_1^+$&621 &56(5) &M1+E2 &1.0$^{+2.0}_{-0.6}$ & { 1.0$^{+0.6}_{-0.7}$}& { 0}   &{ 1.7}  \\	  
\cline{2-9}
         &$ 0_2^+\rightarrow 2_2^+$&323 &35(2) & & & { 100      }& { 100} &{ 100}   \\	  
         &$ 0_2^+\rightarrow 2_1^+$&625 &100(5) & & & { 10.5(8)  }& { 10} &{ 48} \\	  
\cline{2-9}
         &$ 4_2^+\rightarrow 4_1^+$&391 &30(3) &[M1,E2] & & { $<$140   }& { 100} &{ 79} \\	  
         &$ 4_2^+\rightarrow 2_2^+$&531 &100(8) & & & { 100      }& { 100} &{ 100}   \\	  
\cline{2-9}
         &$ 2_3^+\rightarrow 0_2^+$&282 &51(4) & & & { 100      }& { 100} &{ 100}   \\	  
         &$ 2_3^+\rightarrow 3_1^+$&286 &26(2) &(M1)+E2 &$>$5 & { 47(1)    }& { 0}   &{ 21} \\	  
         &$ 2_3^+\rightarrow 4_1^+$&466 &32(2) & & & { 5.1(5)   }& { 75}  &{ 18} \\	  
         &$ 2_3^+\rightarrow 2_2^+$&605 &100(2) &M1+(E2) &$<$0.4 & { $<$0.6   }& { 12}  &{ 9.3}   \\	  
         &$ 2_3^+\rightarrow 2_1^+$&907 &97(2) &E0+(M1,E2) & & { $<$0.6   }& { 0}   &{ 0.03}  \\	  
         &$ 2_3^+\rightarrow 0_1^+$&1203 &14(2) & & & { 0.019(3) }& { 0.04}&{ 0.6}  \\ 
\hline
$^{192}$Pt&$2_3^+\rightarrow 4_1^+$&655  &3.3(4) & & & { 0.85(11)} &{ 64}   &{ 16} \\   
         &$2_3^+\rightarrow 0_2^+$&244  &2.8(3) & & & { 100}      &{ 100}    &{ 100} \\   
         &$2_3^+\rightarrow 3_1^+$&518  &26.8(35) & & & { $<$22}    &{ 0}      &{ 18} \\   
         &$2_3^+\rightarrow 2_2^+$&827  &3.4(3) & & & { $<$0.3}     &{ 9}    &{ 8.4} \\   
         &$2_3^+\rightarrow 2_1^+$&1123  &100(4) & & & { $<$1.7}     &{ 0}      &{ 0.02} \\   
         &$2_3^+\rightarrow 0_1^+$&1439  &4.6(5) & & & { 0.023(3)} &{ 0.1}   &{ 0.7} \\   
\hline 
$^{194}$Pt&$3_1^+\rightarrow 4_1^+$&111  &0.49(15) &[M1,E2] & & { $<$75}  &{ 40}&{ 39}  \\   
      &$3_1^+\rightarrow 2_2^+$&301  &100.0(10) &(M1)+E2 &$>$5 & { 100}  &{ 100}&{ 100}  \\   
      &$3_1^+\rightarrow 2_1^+$&594  &18.4(6) &(M1)+E2 &$>$10 & { $<$0.64}  &{ 0.0}&{ 0.6} \\   
\hline
\end{tabular}
\end{center}
  \caption{See caption of Table \ref{tab-be2-1}.}
  \label{tab-be2-4}
\end{table}

We have compared in Tables \ref{tab-be2-1}, \ref{tab-be2-2}, \ref{tab-be2-3}, and \ref{tab-be2-4} the
experimental relative $B(E2)$ values with the calculated values using the IBM-CM and ECQF.

A first remark concerns the comparison between the calculated values: IBM-CM and ECQF. There appears
an overall agreement on the structure of very strong, intermediate strength and weak relative $B(E2)$
values, which holds for the whole region presented in the extensive table. There are a number of 
cases, such as in $^{184}$Pt (for the decay from the $2^+_4$ and $4^+_3$ levels) and in $^{188}$Pt 
(again decay from the $2^+_4$ level) where large deviations show up. We consider here 121 transitions.

Very much the same conclusion holds when comparing with the experimental relative $B(E2)$ values: for most of the
transitions, the scale of strong, intermediate strength and weak seems to remain intact in comparing
with the theoretical descriptions.

\end{document}